\documentclass[osajnl,preprint,showpacs,superscriptaddress,12pt]{revtex4} 
\usepackage{amsmath,amssymb,graphicx}
\begin{document}
\title{Asymmetrically filled slits in a metal film that split a light beam into two depending on its wavelength}

\author{Danhong Huang}
\author{L. David Wellems}

\affiliation{Air Force Research Laboratory, Space Vehicles Directorate, Kirtland Air Force Base, NM 87117, USA}

\begin{abstract}
By applying a scattering-wave theory, the electromagnetic response of
an arbitrary array of multiple slits perforated on a metallic film and filled with different slit dielectric materials can be studied in an analytical way.
Here, the wavelength-dependent splitting of a light beam into two
by asymmetrically filled slits in a metal film
using intra- and inter-slit dual-wave interferences is fully explored.
We consider a triple-slit structure perforated on a gold film, where the middle slit is used for the surface-plasmon
excitation by a narrow Gaussian beam while the two side slits are used for the detection
of a transmitted surface-plasmon wave propagated from the middle opaque slit either at a particular wavelength
or at double that wavelength, respectively.
For this proposed simple structure, we show that only one of the two side observation slits can be in a passing state for a particular wavelength, but the
other blocked slit will change to a passing state at double that wavelength with a specific design for the slit depth, slit dielectric, and inter-slit distance in the deep sub-wavelength regime.
In this sense, surface-plasmon mediated light transmission becomes wavelength sensitive in our model, and a single light beam can be separated into two according to its wavelength
in the transverse direction parallel to the array. This provides us with a unique way for direct optical reading in the near-field region using a non-spectroscopic approach.
\end{abstract}

\maketitle 

\section{Introduction}
\label{sec1}

Surface-plasmon-polariton modes\,\,\cite{raether} and localized surface plasmons\,\cite{bookg}, which are both localized at an interface between
a bulk conductor and a bulk dielectric, have become very hot research subjects in recent years\,\cite{vidal}.
The extraordinarily high transmission of a $p$-polarized light beam propagating through a two-dimensional honeycomb lattice of holes on a metal film with sub-wavelength diameters\,\cite{ebbesen,ebbesen1,ebbesen2}
depends strongly on the lattice constant and the metal-film thickness (also in the deep sub-wavelength regime)\,\cite{lin,lin2}.
\medskip

On the other hand, studies on the surface-plasmon mediated light transmission by a sub-wavelength structure (including random surface roughness)
on a designed metal surface have also attracted a lot of attention\,\cite{bookaa,maradudin}.
For an optically-opaque metal film on a dielectric substrate, if a single slit is perforated on this film, the excited surface plasmons on the front side of the film can be coupled to the backside by intra-slit
interferences\,\cite{vidal,Perez}. In addition, we demonstrate in this paper that not only the intra-slit interference but also the inter-slit interference for a slit array can
affect the transmission of the excited surface plasmon propagating through the slits, which was not explored in previous works\,\cite{vidal,Perez}.
The inter-slit interference considered in this paper is related only to the surface wave but not to the surface-plasmon wave, since this interference effect
survives even for a perfect electric conductor.
It is also important to mention that the surface-plasmon-polariton mode for a planar surface becomes folded with a finite lattice constant in a periodic array and is split into many branches with a minigap
opened either at the center or at the edge of the first Brillouin zone\,\cite{Baumeier}. Therefore, the surface-plasmon mediated splitting of a light beam studied
in this paper has to consider avoiding these minigaps.
A related work on plasmonic photon sorters for spectral and polarimetric imaging was reported early\,\cite{bull}, which employed coupled bull-eye structures
with a linear modulation in groove depth within each structure.
\medskip

In our previous study\,\cite{huang10}, we have shown the longitudinally color-dependent light focusing by a finite linear array of grooves with various widths in a parabolic pattern,
where various focal lengths of a slit-array aperture were obtained for an incident plane wave with different colors.
Here, rather than using a Green's function formalism\,\cite{Baumeier,Wellems}, we present a scattering-wave theory which utilizes a slit-eigenmode expansion
to treat an arbitrary array of slits having arbitrary spacings, widths and dielectrics.
In addition, the derived scattering-wave theory in this paper provides one with a full description to the surface scattering of light
by removing a so-called ``diagonal'' approximation adopted in a previous related theory\,\cite{vidal,Perez,vidal2}.
Specifically, as an example, in this paper we consider a triple-slit structure in which the middle opaque slit is used for the front-side local surface plasmon
excitation by an incident Gaussian beam either at a particular wavelength or at double that wavelength, and the two side slits are used for the wavelength-dependent detection of
surface-plasmon mediated light-beam splitting in the near-field region. With our designed narrow-slit depth, slit dielectric, and inter-slit distance, we show that one of the two observation slits can be in a passing state
while the other one, at the same time, is in a blocking state for a particular wavelength. Moreover, at double that wavelength, the previously-blocked observation slit switches to a passing state. Therefore,
surface-plasmon mediated light-beam splitting becomes wavelength sensitive in our model, and can be spatially separated in the transverse direction parallel to the array.
As a result, it provides us with a unique way for direct color reading in the near-field region based on a non-spectroscopic technique.
In this paper, we have only cited the most relevant and the most recent advances in the fields of light
scattering and surface plasmons, including works reported by us and other groups. The readers who are interested in details
of this field are referred to the review article by Garcia-Vidal, {\em et al}.\,\cite{vidal}
\medskip

The rest of the paper is organized as follows. In Sec.\,\ref{sec2}, we derive the scattering-wave theory for a non-perfect electric conductor to include the loss of a metallic film in the optical-frequency range by
employing a surface impedance boundary condition. This theory is then applied to study the transmission of an electromagnetic field through an arbitrary  array of slits perforated on the metallic film and
filled with various dielectric materials.
The issue about using the surface impedance boundary condition for a film was extensively discussed in Ref.\,\cite{ao}. In general, the surface impedance boundary condition can be expressed as a linear relation
using a $(2\times 2)$ impedance matrix and takes a nonlocal or an integral form.
The zeroth-order term of the impedance matrix is a local matrix. If the skin depth of a metal film is much smaller than its thickness, which is the situation
to be considered in this paper, the off-diagonal elements of the local impedance matrix, which couple two surfaces of a metal film, can be neglected.
In Sec.\,\ref{sec3}, numerical results are presented to demonstrate both the passing and blocking states of two observation slits at a particular wavelength and at double that wavelength, along with detailed
explanations of these two complementary states based on the intra-slit and inter-slit dual-wave interferences.
The conclusions drawn from these results are briefly summarized in Sec.\,\ref{sec4}.

\section{Scattering-Wave Theory}
\label{sec2}

\subsection{Basic Formalism}
\label{sec2.1}

\begin{figure}[htbp]
\centerline{\includegraphics[width=.8\columnwidth]{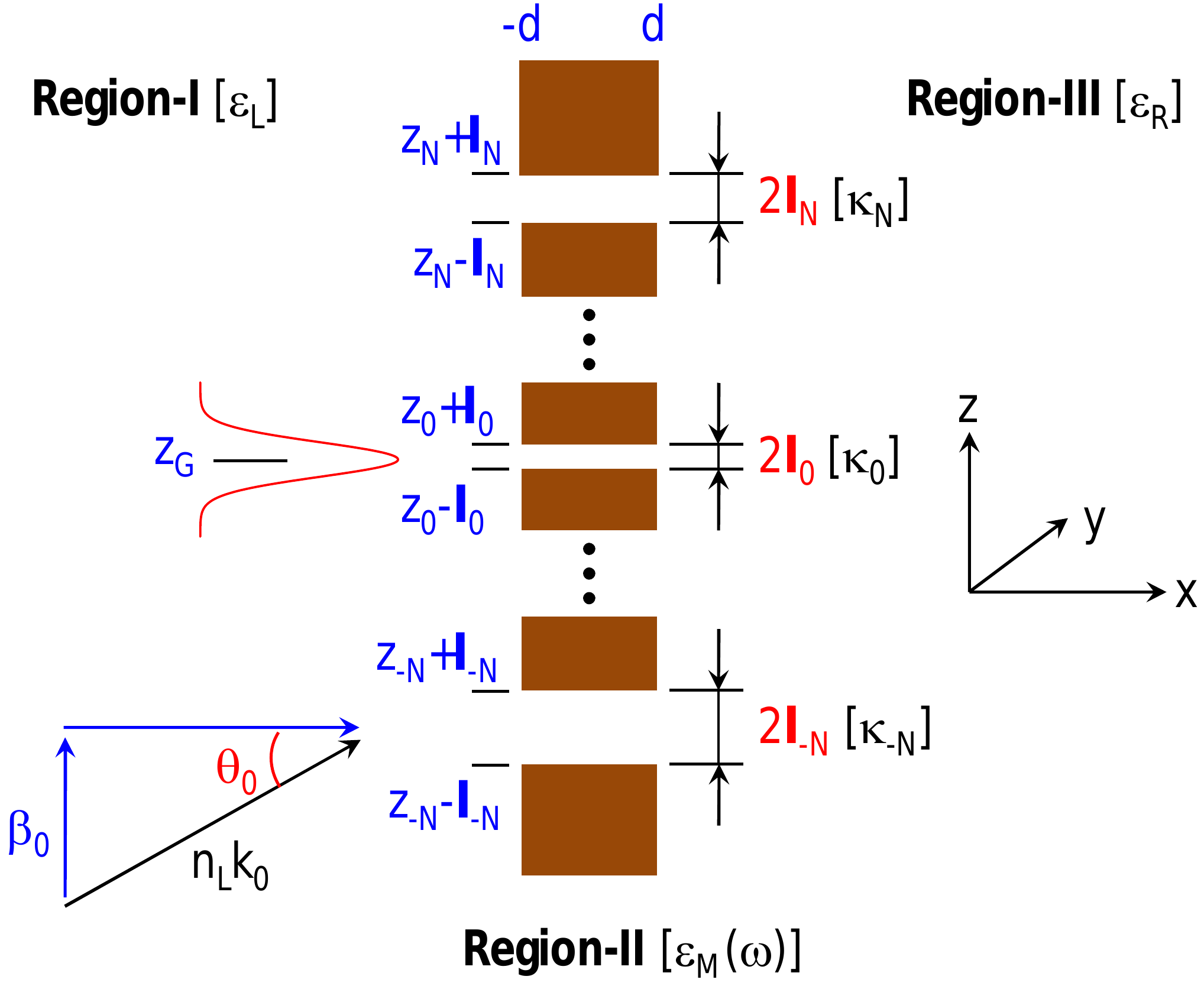}}
\caption{(Color online) Illustration for a $z$ direction slit array
(brown) which extends in the $y$ direction, where $z_j$ and
$2\ell_j$ are the center position and the width of the $j$th slit
with $j=0,\,\pm 1,\,\cdots,\,\pm N$. The regions at the left- and
right-hand side of the slits are denoted as Region I and Region III,
respectively, with real dielectric constants $\epsilon_L$ and
$\epsilon_R$. The region for the slit array is denoted as Region II,
and slits are filled with medium having a dielectric constant
$\kappa_j$ (real or complex) for $j=0,\,\pm 1,\,\cdots,\,\pm N$. The
depth of slits in the $x$ direction is $2d$, and
$\epsilon_M(\omega)$ represents the dielectric function of the metal
film containing slits. A Gaussian beam is incident on the slit array
from the left side with an incident angle $\theta_0$ and at a center
position $z=z_G$. The incident wave number is
$\sqrt{\epsilon_L}\,k_0$ and $\beta_0$ is the incident wave vector
along the $z$ direction.}
\label{f1}
\end{figure}

In this paper, we consider only $p$ polarization for an
electromagnetic (EM) field, written as ${\bf H}=(0,\,H_y,\,0)$, so that
the surface-plasmon (SP) wave on a metal-air interface
can be excited. Here, the ${\bf H}$ field is assumed translationally
invariant in the $y$ direction for the array of slits, shown in
Fig.\,\ref{f1}. We further denote the scalar magnetic-field
amplitude as $u(x,\,z)=H_y(x,\,z)$. The corresponding
electric field can be calculated from ${\bf E}=i/(\omega\epsilon_0\epsilon_s)\,\boldsymbol{\nabla}\times{\bf H}$, where
$\omega$ is the angular frequency and $\epsilon_s$ (real) represents
the relative dielectric constants of the host materials on the left-
($s=L$) and right-hand ($s=R$) side of a metal film. Since the metal film,
which contains a finite slit array, will be treated as a non-perfect
electric conductor (non-PEC), we need to employ the so-called
surface impedance boundary condition\,\cite{maradudin1,Loch,crouse} (SIBC) for the total
EM field. In our current model, the SIBC requires
$\partial u(x,\,z)/\partial x=\eta_s\,u(x,\,z)$ on a metal surface,
where $\eta_s=\pm\,k_0\epsilon_s/[i\sqrt{-\epsilon_{\rm
M}(\omega)}\,]$, $k_0=\omega/c$ is the wave number of the EM field in
vacuum, $\epsilon_{\rm M}(\omega)$ (complex with optical loss) is the metal-film
dielectric function, and the $\pm$ signs refer to the left (minus)
and right (positive) surfaces of the metal film shown in
Fig.\,\ref{f1}. The frequency-dependent
dielectric function $\epsilon_{\rm M}(\omega)$
for the gold film is obtained by interpolation from the data in the paper by Johnson and
Christy\,\cite{christy}.
\medskip

By applying the SIBC on the left-hand side ($L$) of the slit array ($x<-d$), we get

\begin{equation}
\left.\frac{\partial u(x,\,z)}{\partial x}\right|_{x=-d-0}=\left\{
\begin{array}{ll}
\eta_{L}u(x,\,z)|_{x=-d-0} & \ \ \mbox{left surface\ \ $z_j+\ell_j<z<z_{j+1}-\ell_{j+1}$}\\
\frac{\epsilon_L}{\kappa_j}\frac{\partial u(x,\,z)}{\partial x}\left.\right|_{x=-d+0} & \ \ \mbox{middle slit\ \ $|z-z_j|<\ell_j$}\\
\eta_{L}u(x,\,z)|_{x=-d-0} & \ \ \mbox{right surface\ \ $z_{j-1}+\ell_{j-1}<z<z_j-\ell_j$}
\end{array}\right.\ ,
\label{e1}
\end{equation}
where $2d$ is the thickness of the metal film, $j=0,\,\pm 1,\,\pm
2,\,\cdots,\,\pm N$ is the slit index, $z_j$ and $2\ell_j$ are the
center position and the width of the $j$th slit, and $\kappa_j$
(real or complex) is the dielectric constant of the material inside
the $j$th slit. Similarly, after applying the SIBC to the right-hand
side ($R$) of the slit array ($x>d$), we acquire

\begin{equation}
\left.\frac{\partial u(x,\,z)}{\partial x}\right|_{x=d+0}=\left\{
\begin{array}{ll}
\eta_{R}u(x,\,z)|_{x=d+0}  & \ \ \mbox{left surface\ \ $z_j+\ell_j<z<z_{j+1}-\ell_{j+1}$}\\
\frac{\epsilon_R}{\kappa_j}\frac{\partial u(x,\,z)}{\partial x}\left.\right|_{x=d-0} & \ \ \mbox{middle slit\ \ $|z-z_j|<\ell_j$}\\
\eta_{R}u(x,\,z)|_{x=d+0} & \ \ \mbox{right surface\ \ $z_{j-1}+\ell_{j-1}<z<z_j-\ell_j$}
\end{array}\right.\ .
\label{e2}
\end{equation}
If we set $\eta_L=\eta_R=0$ in Eqs.\,(\ref{e1}) and (\ref{e2}), we
will simply get the corresponding boundary conditions for a PEC\,\cite{pec}. In
addition, the continuity of $u(x,\,z)$ is needed for each slit
entry: i.e., $\left. u(x,\,z)\right|_{x=-d-0}=\left.
u(x,\,z)\right|_{x=-d+0}$ and $\left. u(x,\,z)\right|_{x=d-0}=\left.
u(x,\,z)\right|_{x=d+0}$.
For Lamellar metallic gratings, the PEC boundary condition for the slit side walls was used in calculations of plasmon-mediated light reflection, and the results agreed very well with experimental results\cite{just1,just2}.
Therefore, for the PEC the field normal derivative must be
zero along the slit side walls, i.e.,

\begin{equation}
\left.\frac{\partial u(x,\,z)}{\partial z}\right|_{z=z_j\pm\ell_j}=0\ \ \mbox{for all slits\ \ $|x|\leq d$}\ .
\label{e3}
\end{equation}
\medskip

In Region I (the left-hand side of the slit array), the total field, including both the incident and reflected ones, can be written as\,\cite{pec}

\begin{eqnarray}
u^{({\rm I})}(x,\,z)&=&k_0\epsilon_L\int\limits_0^{\infty}
\frac{d\beta}{k_1(\beta)}\,\left[G_s(\beta)\cos(\beta z)
+iG_a(\beta)\sin(\beta z)\right]\,e^{ik_1(\beta)(x + d)}
\nonumber\\
&-&k_0\epsilon_L\int\limits_0^{\infty} \frac{d\beta}{k_1(\beta)}\,\left[A_s(\beta)\,\cos(\beta z)
+iA_a(\beta)\,\sin(\beta z )\right]\,e^{-ik_1(\beta)(x + d)}\ ,
\label{e4}
\end{eqnarray}
where $G_s(\beta)$ and $ G_a(\beta)$ are the symmetric and
anti-symmetric spectral components of an incident Gaussian beam,
given by

\begin{eqnarray}
G_s(\beta)&=&\left[G_n(\beta)+G_p(\beta)\right]\,\cos(\beta z_G)-i\left[G_n(\beta)-G_p(\beta)\right]\,\sin(\beta z_G)\ ,
\nonumber\\
G_a(\beta)&=&\left[G_n(\beta)-G_p(\beta)\right]\,\cos(\beta z_G)-i\left[G_n(\beta)+G_p(\beta)\right]\,\sin(\beta z_G)\ ,
\label{e5}
\end{eqnarray}
$z_G$ dennotes the Gaussian beam center position, and $G_p(\beta)$
and $G_n(\beta)$ in Eq.\,(\ref{e5}) are defined as

\begin{eqnarray}
G_p(\beta)&=&\frac{gk_1(\beta)}{2\sqrt{\pi}\epsilon_L k_0}\,\exp\left[-\frac{g^2(\beta+\beta_0)^2}{4}\right]\,\Theta\left( n_Lk_0-|\beta|\right)\ ,
\nonumber\\
G_n(\beta)&=&\frac{gk_1(\beta)}{2\sqrt{\pi}\epsilon_L k_0}\,\exp\left[-\frac{g^2(\beta-\beta_0)^2}{4}\right]\,\Theta\left(n_Lk_0-|\beta|\right)\ .
\label{e6}
\end{eqnarray}
In Eq.\,(\ref{e6}), $\Theta(\beta)$ is the unit step function, $\beta_0=n_Lk_0\sin\theta_0$, $\theta_0$ is the incident angle of the beam,
$n_L=\sqrt{\epsilon_L}$, and $k_1(\beta)=\sqrt{n_L^2k_0^2-\beta^2}$ can be either real or complex with ${\sf Im}[k_1(\beta)]\geq 0$.
\medskip

In a similar way, we find that in Region III  (the right-hand side
of the slit array) the transmitted field takes the form\,\cite{pec}

\begin{equation}
u^{({\rm III})}(x,\,z)=k_0\epsilon_R \int\limits_0^{\infty}
\frac{d\beta}{k_2(\beta)}\,\left[B_s(\beta)\,\cos(\beta z) +
iB_a(\beta)\,\sin(\beta z)\right]\,e^{ik_2(\beta)(x - d)}\ ,
\label{e7}
\end{equation}
where $n_R=\sqrt{\epsilon_R}$ and
$k_2(\beta)=\sqrt{n_R^2k_0^2-\beta^2}$ with ${\sf Im}[k_2(\beta)]\geq 0$.
\medskip

Finally, in Region II (middle slit array), using the eigenmode expansion [subjected to the boundary condition in Eq.\,(\ref{e3})], we obtain\,\cite{pec}

\begin{eqnarray}
&&u^{({\rm
II})}(x,\,z)=k_0\sum_j\,\Theta(\ell_j-|z-z_j|)\sum_n\,\left\{\frac{\kappa_j}{\sigma_{sn}^j}\left[a_{sn}^j\,e^{i\sigma_{sn}^j(x+d)}-b_{sn}^j\,
e^{-i\sigma_{sn}^j(x-d)}\right]\right.
\nonumber\\
&\times& \cos[\xi_{sn}^j(z-z_j)]+\left.i\frac{\kappa_j}{\sigma_{an}^j}\left[a_{an}^j\,e^{i\sigma_{an}^j(x+d)}-
b_{an}^j\,e^{-i\sigma_{an}^j(x-d)}\right]\,\sin[\xi_{an}^j(z-z_j)]\right\}\
, \label{e8}
\end{eqnarray}
where $n=1,\,2,\,\cdots$ is the eigenmode index,
$\xi_{sn}^j=(\pi/\ell_j)\,(n-1)$ and
$\xi_{an}^j=(\pi/\ell_j)\,(n-1/2)$ are for symmetric and
anti-symmetric slit eigenmodes, respectively, and
$\sigma^j_{sn,\,an}=\sqrt{\kappa_jk_0^2-(\xi_{sn,\,an}^j)^2}$  can be
either real or complex, with ${\sf Re}[\sigma^j_{sn,\,an}]\geq 0$.
\medskip

When $n=0$, the lowest symmetric eigenmode in Eq.\,(\ref{e8}) corresponds to a uniform EM field distribution in the $z$ direction within
each slit. The evanescent waves can exist only when $k_1(\beta)$ or $k_2(\beta)$ is purely imaginary, i.e., $\beta>n_Lk_0$ for the reflection side or $\beta>n_Rk_0$ for the transmission side.
On the other hand, the condition for a pure scattered surface wave is obtained through $k_1(\beta)\rightarrow 0$ or $k_2(\beta)\rightarrow 0$.
This condition can be met by $\beta=n_Lk_0$ for reflection or $\beta=n_Rk_0$ for transmission.

\subsection{SIBC Constraints}

By using the derivative boundary conditions in Eqs.\,(\ref{e1}) and
(\ref{e2}) at $x=\pm\,d$ and the orthogonality of the continuous
Fourier expansions in Eqs.\,(\ref{e4}) and (\ref{e7}), a set of
constraint equations for the unknown Fourier coefficients $A_s(\beta)$,
$A_a(\beta)$, $B_s(\beta)$ and $B_a(\beta)$ can be obtained.
\medskip

At $x=-d$, from Eq.\,(\ref{e1}) we get

\begin{eqnarray}
&& \int\limits_0^{\infty} d\beta\,\left\{\left[G_s(\beta) + A_s(\beta)\right]\,\cos(\beta z)+ i\left[G_a(\beta) +A_a(\beta)\right]\,\sin(\beta z )\right\}
\nonumber\\
&=&-i\eta_L\epsilon_L\int\limits_0^{\infty} \frac{d\beta}{k_1(\beta)}\,\left\{\left[G_s(\beta)-A_s(\beta)\right]\,\cos(\beta z)+
i\left[G_a(\beta)-A_a(\beta)\right]\,\sin(\beta z )\right\}
\label{e9}
\end{eqnarray}
for $z$ values within the non-slit regions of a non-PEC, and

\begin{eqnarray}
&&\int\limits_0^{\infty} d\beta\,\left\{\left[G_s(\beta)+A_s(\beta)\right]\,\cos(\beta z)+ i\left[G_a(\beta)+A_a(\beta)\right]\,\sin(\beta z )\right\}
\nonumber\\
&&=\sum_j\,\Theta(\ell_j-|z-z_j|)\sum_n\,\left[\left(a_{sn}^j +b_{sn}^j\,e^{i\sigma_{sn}^j2d}\right)\,\cos[\xi_{sn}^j(z-z_j)]\right.
\nonumber\\
&&\left.+i \left(a_{an}^j+b_{an}^j\,e^{i\sigma_{an}^j 2d}\right)\,\sin[\xi_{an}^j(z-z_j)]\right]
\label{e10}
\end{eqnarray}
for the slit regions. Similarly, at $x=d$ we find from Eq.\,(\ref{e2})

\begin{eqnarray}
&&\int\limits_0^{\infty} d\beta\,\left[B_s(\beta)\,\cos(\beta z)+ iB_a(\beta)\,\sin(\beta z )\right]
\nonumber\\
&=&-i\eta_R\epsilon_R\int\limits_0^{\infty} \frac{d\beta}{k_2(\beta)}\,\left[B_s(\beta)\,\cos(\beta z)+
i B_a(\beta)\,\sin(\beta z )\right]
\label{e11}
\end{eqnarray}
for $z$ values within the non-slit regions, and

\begin{eqnarray}
&&\int\limits_0^{\infty} d\beta\,\left[B_s(\beta)\,\cos(\beta z)+ iB_a(\beta)\,\sin(\beta z )\right]=\sum_j\,\Theta(\ell_j-|z-z_j|)
\sum_n\,\left[\left(b_{sn}^j \right.\right.
\nonumber\\
&&\left.\left.+a_{sn}^j\,e^{i\sigma_{sn}^j 2d}\right)\cos[\xi_{sn}^j(z-z_j)]+i\left(b_{an}^j +  a_{an}^j\,e^{i\sigma_{an}^j 2 d}\right)\,\sin[\xi_{an}^j(z-z_j)]\right]
\label{e12}
\end{eqnarray}
for the slit regions.

\subsection{Projection of SIBC}

Since the combination of Eqs.\,(\ref{e9}) and (\ref{e10}) extends
over the left surface of a metal film, we can project out the
symmetric and anti-symmetric Fourier coefficients in
$u^{({\rm I})}(x,\,z)$ through multiplying these two equations by
$\cos(\beta^\prime z)$ or $\sin(\beta^\prime z)$ and integrating
over $z$ afterwards. This leads to

\begin{eqnarray}
&&A_s(\beta) +G_s(\beta)
=\sum_n\,\left\{\sum_j\,\frac{\ell_j}{\pi}\,\left[\left(a_{sn}^j+b_{sn}^j\,e^{2i\sigma_{sn}^jd}\right)\,Q_{sn}^j(\beta)\,\cos(\beta z_j)\right.\right.
\nonumber\\
&-&i\left.\left.\left(a_{an}^j+ b_{an}^j\,e^{2i\sigma_{an}^jd}\right)\,Q_{an}^j(\beta)\,\sin(\beta z_j)\right]\right\}
\nonumber\\
&-&i\eta_L \int\limits_0^\infty \frac{d\beta^\prime}{k_1(\beta^\prime)}\,\left[P_s(\beta,\,\beta^\prime)+ W_s(\beta,\,\beta^\prime)\right]\,
\left[G_s(\beta^\prime) - A_s(\beta^\prime)\right]
\nonumber\\
&+&\eta_L \int\limits_0^\infty \frac{d\beta^\prime}{k_1(\beta^\prime)}\,\left[P_c(\beta,\,\beta^\prime)+W_c(\beta,\,\beta^\prime)\right]\,
\left[G_a(\beta^\prime) -  A_a(\beta^\prime)\right]\ ,
\label{e13}
\end{eqnarray}

\begin{eqnarray}
&&A_a(\beta) +G_a(\beta)
=\sum_n\,\left\{\sum_j\,\frac{\ell_j}{\pi}\,\left[-i\left(a_{sn}^j+b_{sn}^j\,e^{2i\sigma_{sn}^jd}\right)\,Q_{sn}^j(\beta)\,\sin(\beta z_j)\right.\right.
\nonumber\\
&+&\left.\left.\left(a_{an}^j+ b_{an}^j\,e^{2i\sigma_{an}^jd}\right)\,Q_{an}^j(\beta)\,\cos(\beta z_j)\right]\right\}
\nonumber\\
&-&\eta_L \int\limits_0^\infty \frac{d\beta^\prime}{k_1(\beta^\prime)}\,\left[P_{c}(\beta^\prime,\,\beta)+W_{c}(\beta^\prime,\,\beta)\right]\,
\left[G_s(\beta^\prime) - A_s(\beta^\prime)\right]
\nonumber\\
&-&i\eta_L \int\limits_0^\infty \frac{d\beta^\prime}{k_1(\beta^\prime)}\,\left[P_{a}(\beta,\,\beta^\prime)+W_{a}(\beta,\,\beta^\prime)\right]\,
\left[G_a(\beta^\prime) -  A_a(\beta^\prime)\right]\ ,
\label{e14}
\end{eqnarray}
where the definitions of $Q_{sn}^j(\beta)$ and $Q_{an}^j(\beta)$ can be found from Appendix A.
\medskip

For the same reason, using Eqs.\,(\ref{e11}) and
(\ref{e12}) we can also project out the symmetric and
anti-symmetric Fourier coefficients in $u^{({\rm III})}(x,\,z)$
through multiplying them by $\cos(\beta^\prime z)$ or $\sin(\beta^\prime z)$ and
doing a follow-up $z$ integration. This yields

\begin{eqnarray}
B_s(\beta)
&=&\sum_n\,\left\{\sum_j\,\frac{\ell_j}{\pi}\,\left[\left(b_{sn}^j+a_{sn}^j\,e^{2i\sigma_{sn}^jd}\right)\,Q_{sn}^j(\beta)\,\cos(\beta z_j)\right.\right.
\nonumber\\
&-&\left.\left.i\left(b_{an}^j+ a_{an}^j\,e^{2i\sigma_{an}^jd}\right)\,Q_{an}^j(\beta)\,\sin(\beta z_j)\right]\right\} \
\nonumber\\
&-&i\eta_R \int\limits_0^\infty \frac{d\beta^\prime}{k_2(\beta^\prime)}\,\left[P_s(\beta,\,\beta^\prime)+ W_s(\beta,\,\beta^\prime)\right]\,B_s(\beta^\prime)
\nonumber\\
&+&\eta_R \int\limits_0^\infty \frac{d\beta^\prime}{k_2(\beta^\prime)}\,\left[P_c(\beta,\,\beta^\prime)+W_c(\beta,\,\beta^\prime)\right]\,B_a(\beta^\prime)\ ,
\label{e17}
\end{eqnarray}

\begin{eqnarray}
B_a(\beta)
&=&\sum_n\,\left\{\sum_j\,\frac{\ell_j}{\pi}\,\left[-i\left(a_{sn}^j+b_{sn}^j\,e^{2i\sigma_{sn}^jd}\right)\,Q_{sn}^j(\beta)\,\sin(\beta z_j)\right.\right.
\nonumber\\
&+&\left.\left.\left(a_{an}^j+ b_{an}^j\,e^{2i\sigma_{an}^jd}\right)\,Q_{an}^j(\beta)\,\cos(\beta z_j)\right]\right\} \
\nonumber\\
&-&\eta_R \int\limits_0^\infty \frac{d\beta^\prime}{k_2(\beta^\prime)}\,\left[P_{c}(\beta^\prime,\,\beta)+W_{c}(\beta^\prime,\,\beta)\right]\,B_s(\beta^\prime)
\nonumber\\
&-&i\eta_R \int\limits_0^\infty \frac{d\beta^\prime}{k_2(\beta^\prime)}\,\left[P_{a}(\beta,\,\beta^\prime) +W_{a}(\beta,\,\beta^\prime)\right]\,B_a(\beta^\prime)\ .
\label{e18}
\end{eqnarray}
The definition of the coupling matrices $P_s(\beta,\,\beta^\prime)$, $P_a(\beta,\,\beta^\prime)$, $P_c(\beta,\,\beta^\prime)$, $W_s(\beta,\,\beta^\prime)$, $W_a(\beta,\,\beta^\prime)$, and $W_c(\beta,\,\beta^\prime)$ can be found in Appendix B.
Without loss of generality, we assume an order for the slit array
$-\infty<(z_{-N}-\ell_{-N})<(z_{-N}+\ell_{-N})<\cdots<(z_{-1}-\ell_{-1})<(z_{-1}+\ell_{-1})<(z_{0}-\ell_{0})<
(z_0+\ell_0)<(z_1-\ell_1)<(z_1+\ell_1)<\cdots<(z_N-\ell_N)\leq(z_N+\ell_N)<\infty$.
\medskip

If $\eta_L=\eta_R=0$ for a PEC, we can explicitly express\,\cite{pec} the
continuous Fourier expansion coefficients $A_s(\beta)$,
$A_a(\beta)$, $B_s(\beta)$ and $B_a(\beta)$ by the discrete Fourier
expansion coefficients $a_{sn}^j$, $b_{sn}^j$, $a_{an}^j$ and
$b_{an}^j$ through Eqs.\,(\ref{e13}), (\ref{e14}), (\ref{e17}) and
(\ref{e18}). If there exists only one slit ($j=0$ and $z_0=0$), the PEC single slit
does not couple symmetric modes to anti-symmetric ones\,\cite{Serdyuk}.
In addition, we find
$W_s(\beta,\,\beta^\prime)=W_a(\beta,\,\beta^\prime)=W_c(\beta,\,\beta^\prime)=0$
and $P_c(\beta,\,\beta^\prime)=0$ in this case. Therefore, the
non-PEC single slit cannot couple symmetric modes to anti-symmetric ones.
For a symmetric distribution of slits with respect to
$z_0=0$, we always have $|z_{-N}-\ell-{_N}|=z_N+\ell_N$, leading to $P_c(\beta,\,\beta^\prime)=0$. However, in this
multi-slit case, $W_c(\beta,\,\beta^\prime)\neq 0$. As a result,
symmetric and anti-symmetric modes are coupled to each other.
If we treat the SIBC in Eqs.\,(\ref{e13}), (\ref{e14}), (\ref{e17}) and (\ref{e18})
as a perturbation for small $\eta_L$ and $\eta_R$, we can express $A_s(\beta)$,
$A_a(\beta)$, $B_s(\beta)$ and $B_a(\beta)$ by the discrete Fourier
expansion coefficients $a_{sn}^j$, $b_{sn}^j$, $a_{an}^j$ and
$b_{an}^j$ to the leading order of the perturbation.
Alternatively, if only the diagonal contributions for the term proportional to $\eta_L$ or $\eta_R$ are kept\,\cite{vidal,Perez,vidal2} in
Eqs.\,(\ref{e13}), (\ref{e14}), (\ref{e17}) and (\ref{e18})
[i.e., including only the terms with $\beta^\prime=\pm\beta$ in the integrals], $A_s(\beta)$,
$A_a(\beta)$, $B_s(\beta)$ and $B_a(\beta)$ can also be expressed by the discrete Fourier
expansion coefficients $a_{sn}^j$, $b_{sn}^j$, $a_{an}^j$ and
$b_{an}^j$. However, such a simplification\,\cite{vidal,Perez,vidal2} needs to be justified physically.
\medskip

The projected SBIC in Eqs.\,(\ref{e13})-(\ref{e18}) leads to the approximate field equations
derived in References \cite{vidal,Perez,vidal2} after applying the ``diagonal approximation'' for simplification.
It is clear from Eqs.\,(\ref{e13})-(\ref{e18}) that such a ``diagonal approximation'' can only be justified when the coefficients $P_s(\beta.\,\beta^\prime)$, $P_c(\beta.\,\beta^\prime)$, $P_a(\beta.\,\beta^\prime)$,
as well as the coefficients $W_s(\beta.\,\beta^\prime)$, $W_c(\beta.\,\beta^\prime)$, $W_a(\beta.\,\beta^\prime)$, are either peaked around $\beta=\beta^\prime$
or negligibly small. We further realize from Appendices B \& C
that $P_s(\beta.\,\beta^\prime)$, $P_a(\beta.\,\beta^\prime)$ always peak around $\beta=\beta^\prime$ due to the existence of the $\delta(\beta-\beta^\prime)$ terms;
$W_s(\beta.\,\beta^\prime)$, $W_a(\beta.\,\beta^\prime)$ can peak around $\beta=\beta^\prime$ only if neighboring slits are well separated from each other in comparison with the inverses of $\beta$ and $\beta^\prime$
due to the existence of sinc-function terms;
both $P_c(\beta.\,\beta^\prime)$ and $W_c(\beta.\,\beta^\prime)$ can be neglected only for very large $\beta,\,\beta^\prime$ and $\beta\neq\beta^\prime$ due to the existence of cosine terms.
Therefore, the advantage of the current theory is its immunization from such restrictions and it can be applied to general cases.

\subsection{Integral Equations}

In order to get the closed-form integral equations for a non-PEC, we
look to express $a_{sn}^j$, $a_{an}^j$, $b_{sn}^j$ and
$b_{an}^j$ in Eqs.\,(\ref{e13}), (\ref{e14}), (\ref{e17}) and
(\ref{e18}) by $A_s(\beta)$, $A_a(\beta)$, $B_s(\beta)$ and
$B_a(\beta)$. This can be achieved by using the orthogonality
of the discrete Fourier expansion in Eq.\,(\ref{e8}) and multiplying
both sides of the continuity conditions for slit entries and exits, i.e.
$u^{({\rm I})}(x=-d,\,z)=u^{({\rm II})}(x=-d,\,z)$ and $u^{({\rm
III})}(x=d,\,z)=u^{({\rm II})}(x=d,\,z)$, by
$\cos[\xi_{sn^\prime}^{j^\prime}(z-z_{j^\prime})]$ or
$\sin[\xi_{an^\prime}^{j^\prime}(z-z_{j^\prime})]$, which is
followed by an integration of $z$ over all the slit regions. This
yields

\begin{eqnarray}
&&\frac{\chi_n \kappa_j}{\sigma_{sn}^j \epsilon_L}\,\left(a_{sn}^j-
b_{sn}^j\,e^{2i\sigma_{sn}^jd}\right)=\int\limits_{0}^{\infty}
\frac{d\beta}{k_1(\beta)}\,
\left[G_s(\beta)-A_s(\beta)\right]\,Q_{sn}^j(\beta)\,\cos(\beta z_j)
\nonumber\\
&+&i\int\limits_{0}^{\infty} \frac{d\beta}{k_1(\beta)}\,
\left[G_a(\beta)-A_a(\beta)\right]\,Q_{sn}^j(\beta)\,\sin(\beta z_j)\ ,
\label{e25}
\end{eqnarray}

\begin{eqnarray}
&&\frac{\chi_n \kappa_j}{\sigma_{an}^j \epsilon_L}\,\left(a_{an}^j-
b_{an}^j\,e^{2i\sigma_{an}^jd}\right)=i\int\limits_{0}^{\infty}
\frac{d\beta}{k_1(\beta)}\,
\left[G_s(\beta)-A_s(\beta)\right]\,Q_{an}^j(\beta)\,\sin(\beta z_j)
\nonumber\\
&+&\int\limits_{0}^{\infty} \frac{d\beta}{k_1(\beta)}\,
\left[G_a(\beta)-A_a(\beta)\right]\,Q_{an}^j(\beta)\,\cos(\beta
z_j)\ , \label{e26}
\end{eqnarray}

\begin{eqnarray}
&&\frac{\chi_n \kappa_j}{\sigma_{sn}^j
\epsilon_R}\,\left(a_{sn}^j\,e^{2i\sigma_{sn}^jd} -
b_{sn}^j\right)=\int\limits_{0}^{\infty}
\frac{d\beta}{k_2(\beta)}\, B_s(\beta)\,Q_{sn}^j(\beta)\,\cos(\beta
z_j)
\nonumber\\
&+&i\int\limits_{0}^{\infty} \frac{d\beta}{k_2(\beta)}\,
B_a(\beta)\,Q_{sn}^j(\beta)\,\sin(\beta z_j)\ ,
\label{e27}
\end{eqnarray}

\begin{eqnarray}
&&\frac{\chi_n \kappa_j}{\sigma_{an}^j \epsilon_R}\,\left(a_{an}^j
e^{2i\sigma_{an}^jd}- b_{an}^j\right)=i\int\limits_{0}^{\infty}
\frac{d\beta}{k_2(\beta)}\, B_s(\beta)\,Q_{an}^j(\beta)\,\sin(\beta
z_j)
\nonumber\\
&+&\int\limits_{0}^{\infty} \frac{d\beta}{k_2(\beta)}\,
B_a(\beta)\,Q_{an}^j(\beta)\,\cos(\beta z_j)\ ,
\label{e28}
\end{eqnarray}
where $\chi_n=2$ for $n=1$ and $\chi_n=1$ for $n\neq 1$.
Equations\ (\ref{e25}) and (\ref{e26}) connect the $j$th-slit field at the entry edge to the total field in Region I,
where the forward-moving amplitude, $a_{sn}^j$ or $a_{an}^j$, and backward-moving amplitude, $b_{sn}^j$ or $b_{an}^j$, can be viewed as two independent
interfering waves with phase delays of $2i\sigma_{sn}^jd$ or $2i\sigma_{an}^jd$.
The same arguments can be applied to Eqs.\,(\ref{e27}) and (\ref{e28}) at the exit edge of the $j$th slit.
\medskip

Equations\ (\ref{e25}) through Eq.\,(\ref{e28}) can be formally solved
analytically, which leads to

\begin{eqnarray}
a_{sn}^j=\frac{Y^{(2)}_{jn}-e^{-2i\sigma_{sn}^jd}\,Y^{(1)}_{jn}}{2i\sin(2\sigma_{sn}^jd)}\
,\ \ \ \ \ \ \ \ b_{sn}^j=\frac{e^{-2i\sigma_{sn}^jd}\,Y^{(2)}_{jn} -
Y^{(1)}_{jn} }{2i\sin(2\sigma_{sn}^jd)}\ , \nonumber
\\
\nonumber\\
a_{an}^j=\frac{X^{(2)}_{jn}-e^{-2i\sigma_{an}^jd}\,X^{(1)}_{jn}}{2i\sin(2\sigma_{an}^jd)}\
,\ \ \ \ \ \ \ \ b_{an}^j=\frac{e^{-2i\sigma_{an}^jd}\,X^{(2)}_{jn} -
X^{(1)}_{jn} }{2i\sin(2\sigma_{an}^jd)}\ . \label{e29}
\end{eqnarray}
Here, $a_{sn}^j$ and $a_{an}^j$ represents the forward-moving waves,
while $b_{sn}^j$ and $b_{an}^j$ represents the backward-moving waves.
Whenever $2\sigma_{sn}^jd/\pi$ or $2\sigma_{an}^jd/\pi$ becomes an
integer, the constructive or destructive dual-wave interference will occur at two the edges of
the $j$th slit for the $n$th eigenmode. In this case, the slit behaves either like a passing filter or like a cavity for field trapping.
The definitions of $Y^{(1)}_{jn}$, $Y^{(2)}_{jn}$, $X^{(1)}_{jn}$, and $X^{(2)}_{jn}$ introduced in Eq.\,(\ref{e29}) can be found in Appendix C.
Here, $X^{(1)}_{jn}$ and $Y^{(1)}_{jn}$ come from the contribution
of $u^{({\rm I})}(x,\,z)$ at the left slit entry, while
$X^{(2)}_{jn}$ and $Y^{(2)}_{jn}$ come from the contribution of
$u^{({\rm III})}(x,\,z)$ at the right slit exit. For a PEC, by
substituting Eqs.\,(\ref{e13}), (\ref{e14}), (\ref{e17}) and
(\ref{e18}) with $\eta_L=\eta_R=0$ into Eqs.\,(\ref{e25})--(\ref{e28}), we get a
set of inhomogeneous linear equations\,\cite{pec} with respect to $a_{sn}^j$,
$b_{sn}^j$, $a_{an}^j$ and $b_{an}^j$.
\medskip

Using the results in Eq.\,(\ref{e29}), from Eqs.\,(\ref{e13}), (\ref{e14}), (\ref{e17}) and (\ref{e18}) we finally obtain
four integral equations for $A_s(\beta)$, $A_a(\beta)$,
$B_s(\beta)$ and $B_a(\beta)$

\begin{eqnarray}
&&G_s(\beta)+A_s(\beta)=-i\eta_L\int\limits_0^\infty \frac{d\beta^\prime}{k_1(\beta^\prime)}\,\left[P_{s}(\beta,\,\beta^\prime)+W_{s}(\beta,\,\beta^\prime)\right]\,\left[ G_s(\beta^\prime)-A_s(\beta^\prime)\right]
\nonumber\\
&+&\eta_L\int\limits_0^\infty \frac{d\beta^\prime}{k_1(\beta^\prime)}\,\left[P_{c}(\beta,\,\beta^\prime)+W_{c}(\beta,\,\beta^\prime)\right]\,\left[G_a(\beta^\prime)-A_a(\beta^\prime)\right]
\nonumber\\
&+&\int\limits_0^\infty \frac{d\beta^\prime}{k_2(\beta^\prime)}\,\left[T_s^{(1)}(\beta,\,\beta^\prime)\,B_s(\beta^\prime)+iT_s^{(2)}(\beta,\,\beta^\prime)\,B_a(\beta^\prime)\right]
\nonumber\\
&-&\int\limits_0^\infty \frac{d\beta^\prime}{k_1(\beta^\prime)}\,\left\{T_s^{(3)}(\beta,\,\beta^\prime)\,\left[G_s(\beta^\prime)-A_s(\beta^\prime)\right]+iT_s^{(4)}(\beta,\,\beta^\prime)\,
\left[G_a(\beta^\prime)-A_a(\beta^\prime)\right]\right\}
\nonumber\\
&-&i\int\limits_0^\infty \frac{d\beta^\prime}{k_2(\beta^\prime)}\,\left[R_s^{(1)}(\beta,\,\beta^\prime)\,B_a(\beta^\prime)+iR_s^{(2)}(\beta,\,\beta^\prime)\,B_s(\beta^\prime)\right]
\nonumber\\
&+&i\int\limits_0^\infty \frac{d\beta^\prime}{k_1(\beta^\prime)}\,\left\{R_s^{(3)}(\beta,\,\beta^\prime)\,\left[G_a(\beta^\prime)-A_a(\beta^\prime)\right]\right.
\nonumber\\
&+&\left.iR_s^{(4)}(\beta,\,\beta^\prime)\,\left[G_s(\beta^\prime)-A_s(\beta^\prime)\right]\right\}\ ,
\label{e34}
\end{eqnarray}

\begin{eqnarray}
&&G_a(\beta)+A_a(\beta)=-\eta_L\int\limits_0^\infty
\frac{d\beta^\prime}{k_1(\beta^\prime)}\,\left[P_{c}(\beta^\prime,\,\beta)+W_{c}(\beta^\prime,\,\beta)\right]\,\left[
G_s(\beta^\prime)-A_s(\beta^\prime)\right]
\nonumber\\
&-&i\eta_L\int\limits_0^\infty
\frac{d\beta^\prime}{k_1(\beta^\prime)}\,\left[P_{a}(\beta,\,\beta^\prime)+W_{a}(\beta,\,\beta^\prime)\right]\,\left[G_a(\beta^\prime)-A_a(\beta^\prime)\right]
\nonumber\\
&-&i\int\limits_0^\infty
\frac{d\beta^\prime}{k_2(\beta^\prime)}\,\left[T_a^{(1)}(\beta,\,\beta^\prime)\,B_s(\beta^\prime)+iT_a^{(2)}(\beta,\,\beta^\prime)\,B_a(\beta^\prime)\right]
\nonumber\\
&+&i\int\limits_0^\infty
\frac{d\beta^\prime}{k_1(\beta^\prime)}\,\left\{T_a^{(3)}(\beta,\,\beta^\prime)\,\left[G_s(\beta^\prime)-A_s(\beta^\prime)\right]+iT_a^{(4)}(\beta,\,\beta^\prime)\,
\left[G_a(\beta^\prime)-A_a(\beta^\prime)\right]\right\}
\nonumber\\
&+&\int\limits_0^\infty
\frac{d\beta^\prime}{k_2(\beta^\prime)}\,\left[R_a^{(1)}(\beta,\,\beta^\prime)\,B_a(\beta^\prime)+iR_a^{(2)}(\beta,\,\beta^\prime)\,B_s(\beta^\prime)\right]
\nonumber\\
&-&\int\limits_0^\infty
\frac{d\beta^\prime}{k_1(\beta^\prime)}\,\left\{R_a^{(3)}(\beta,\,\beta^\prime)\,\left[G_a(\beta^\prime)-A_a(\beta^\prime)\right]\right.
\nonumber\\
&+&\left.iR_a^{(4)}(\beta,\,\beta^\prime)\,\left[G_s(\beta^\prime)-A_s(\beta^\prime)\right]\right\}\
, \label{e35}
\end{eqnarray}

\begin{eqnarray}
&&B_s(\beta)=-i\eta_R\int\limits_0^\infty\frac{d\beta^\prime}{k_2(\beta^\prime)}\,\left[P_{s}(\beta,\,\beta^\prime)+W_{s}(\beta,\,\beta^\prime)\right]\,B_s(\beta^\prime)
\nonumber\\
&+&\eta_R\int\limits_0^\infty\frac{d\beta^\prime}{k_2(\beta^\prime)}\,\left[P_{c}(\beta,\,\beta^\prime)+W_{c}(\beta,\,\beta^\prime)\right]\,B_a(\beta^\prime)
\nonumber\\
&+&\int\limits_0^\infty\frac{d\beta^\prime}{k_2(\beta^\prime)}\,\left[N_s^{(1)}(\beta,\,\beta^\prime)\,B_s(\beta^\prime)+iN_s^{(2)}(\beta,\,\beta^\prime)\,B_a(\beta^\prime)\right]
\nonumber\\
&-&\int\limits_0^\infty \frac{d\beta^\prime}{k_1(\beta^\prime)}\,\left\{N_s^{(3)}(\beta,\,\beta^\prime)\,\left[G_s(\beta^\prime)-A_s(\beta^\prime)\right]+iN_s^{(4)}(\beta,\,\beta^\prime)
\left[G_a(\beta^\prime)-A_a(\beta^\prime)\right]\right\}
\nonumber\\
&-&i\int\limits_0^\infty\frac{d\beta^\prime}{k_2(\beta^\prime)}\,\left[M_s^{(1)}(\beta,\,\beta^\prime)\,B_a(\beta^\prime)+iM_s^{(2)}(\beta,\,\beta^\prime)\,B_s(\beta^\prime)\right]
\nonumber\\
&+&i\int\limits_0^\infty\frac{d\beta^\prime}{k_1(\beta^\prime)}\,\left\{M_s^{(3)}(\beta,\,\beta^\prime)\,\left[G_a(\beta^\prime)-A_a(\beta^\prime)\right]\right.
\nonumber\\
&+&\left.iM_s^{(4)}(\beta,\,\beta^\prime)\,
\left[G_s(\beta^\prime)-A_s(\beta^\prime)\right]\right\}\ ,
\label{e36}
\end{eqnarray}

\begin{eqnarray}
&&B_a(\beta)=-\eta_R\int\limits_0^\infty\frac{d\beta^\prime}{k_2(\beta^\prime)}\,\left[P_{c}(\beta^\prime,\,\beta)+W_{c}(\beta^\prime,\,\beta)\right]\,B_s(\beta^\prime)
\nonumber\\
&-&i\eta_R\int\limits_0^\infty\frac{d\beta^\prime}{k_2(\beta^\prime)}\,\left[P_{a}(\beta,\,\beta^\prime)+W_{a}(\beta,\,\beta^\prime)\right]\,B_a(\beta^\prime)
\nonumber\\
&-&i\int\limits_0^\infty\frac{d\beta^\prime}{k_2(\beta^\prime)}\,\left[N_a^{(1)}(\beta,\,\beta^\prime)\,B_s(\beta^\prime)+iN_a^{(2)}(\beta,\,\beta^\prime)\,B_a(\beta^\prime)\right]
\nonumber\\
&+&i\int\limits_0^\infty
\frac{d\beta^\prime}{k_1(\beta^\prime)}\,\left\{N_a^{(3)}(\beta,\,\beta^\prime)\,\left[G_s(\beta^\prime)-A_s(\beta^\prime)\right]+iN_a^{(4)}(\beta,\,\beta^\prime)
\left[G_a(\beta^\prime)-A_a(\beta^\prime)\right]\right\}
\nonumber\\
&+&\int\limits_0^\infty\frac{d\beta^\prime}{k_2(\beta^\prime)}\,\left[M_a^{(1)}(\beta,\,\beta^\prime)\,B_a(\beta^\prime)+iM_a^{(2)}(\beta,\,\beta^\prime)\,B_s(\beta^\prime)\right]
\nonumber\\
&-&\int\limits_0^\infty\frac{d\beta^\prime}{k_1(\beta^\prime)}\,\left\{M_a^{(3)}(\beta,\,\beta^\prime)\,\left[G_a(\beta^\prime)-A_a(\beta^\prime)\right]\right.
\nonumber\\
&+&\left.iM_a^{(4)}(\beta,\,\beta^\prime)\,
\left[G_s(\beta^\prime)-A_s(\beta^\prime)\right]\right\}\ .
\label{e37}
\end{eqnarray}
The expressions for the matrices
$T_{s}^{(\alpha)}(\beta,\,\beta^\prime)$,
$R_{s}^{(\alpha)}(\beta,\,\beta^\prime)$,
$M_{s}^{(\alpha)}(\beta,\,\beta^\prime)$ and
$N_{s}^{(\alpha)}(\beta,\,\beta^\prime)$, as well as for
$T_{a}^{(\alpha)}(\beta,\,\beta^\prime)$,
$R_{a}^{(\alpha)}(\beta,\,\beta^\prime)$,
$M_{a}^{(\alpha)}(\beta,\,\beta^\prime)$ and
$N_{a}^{(\alpha)}(\beta,\,\beta^\prime)$, with $\alpha=1,\,2,\,3,\,4$
are given explicitly in Appendix\ D. In general,
$T_{s,\,a}^{(1,\,2)}(\beta,\,\beta^\prime)$,
$R_{s,\,a}^{(1,\,2)}(\beta,\,\beta^\prime)$,
$M_{s,\,a}^{(1,\,2)}(\beta,\,\beta^\prime)$ and
$N_{s,\,a}^{(1,\,2)}(\beta,\,\beta^\prime)$ come from the
contribution of $u^{({\rm III})}(x,\,z)$, while
$T_{s,\,a}^{(3,\,4)}(\beta,\,\beta^\prime)$,
$R_{s,\,a}^{(3,\,4)}(\beta,\,\beta^\prime)$,
$M_{s,\,a}^{(3,\,4)}(\beta,\,\beta^\prime)$ and
$N_{s,\,a}^{(3,\,4)}(\beta,\,\beta^\prime)$ come from the
contribution of $u^{({\rm I})}(x,\,z)$. If the array is infinite and all the slits, as well as the filled dielectric materials, are identical for a periodic system,\,\cite{Loch,crouse,huang1,huang2},
the continuous variable $\beta$ becomes
a discrete reciprocal wave number $j(2\pi/{\cal D})$ where ${\cal D}$ is the array period and $j=0,\,\pm 1,\,\cdots$.
\medskip

There are three different interferences discussed in this paper, namely, intra-slit and inter-slit dual-wave interferences and Fabry-P\'erot sole-wave interference. First, for the intra-slit dual-wave interference, we consider the interference between the forward and backward moving slit waves. For the backward moving wave, there already exists a $\pi$ phase delay and the latter interferes with the former after its reflection from the entrance side of a slit.
Therefore, the constructive interference condition (on-state) between these two waves is simply given by $\sqrt{\kappa}\,k_04d=(2m-1)\pi$, where $k_0$ is the wave number in vacuum, $\kappa$ is the slit dielectric constant and $m=\pm 1,\,\pm 2,\,\cdots$ is an integer.For the same reason, the destructive interference condition (off-state) is given by $\sqrt{\kappa}\,k_04d=2m\pi$. Second, by the inter-slit dual-wave interference we mean the interference between a backward moving reflected slit wave and a surface wave propagated from another slit and entering forward into the side slit studied.In comparison with the intra-slit dual-wave interference, there exists an additional phase compensation for the forward moving slit wave in this case, which depends only on the slit separation for the surface wave but not on the metal dielectric function. Therefore, we expect a strong influence from the inter-slit dual-wave interference on the intra-slit dual-wave interference. At last, for the Fabry-P\'erot sole-wave interference, on the other hand, it is the interference between a forward moving slit wave and the same wave after it has been reflected twice successively by the exit and entrance sides of a slit. Therefore, the constructive Fabry-P\'erot interference condition takes the form of $\sqrt{\kappa}\,k_04d=2m\pi$, which sits at the same position as that of the destructive intra-slit dual-wave interference, and its strength goes up with increased finesse in a longer slit.

\section{Numerical Results and Discussions}
\label{sec3}

In this paper, we would like to demonstrate a direct optical reading of fundamental and second-harmonic near-field photon emissions in the near-field region
using a non-spectroscopic technique with specifically designed slit depth, slit dielectric material, and inter-slit distance.
In our calculations, we take $\epsilon_L=1$ (air) and $\epsilon_R=20,25$ (high-dielectric-constant oxides).
For a triple-slit structure, we illuminate the middle slit and find conditions under which one of the side slits is in a passing state for a particular wavelength (with the other slit in a blocking state)
while the other side slit is in a passing state for twice that wavelength (with the other slit again in a blocking state).
Here, the passing state of a slit refers to the fact that a surface wave propagated from the middle slit
can pass through a vertical slit to get to Region III
from Region I, while the blocking state corresponds to a surface wave not passing through a slit due to interference effects.
From Eqs.\,(\ref{e25})--(\ref{e28}) we find that the forward and backward moving waves inside a slit can be treated as two independent waves, which are decided
by the fields in Regions I and III, respectively. As a result, the intra-slit interference in this paper can be viewed as a dual-wave interference,
which is quite different from the geometric series result of a typical Fabry-P\'erot (FP) slit model.
FP cavity analysis utilizes intra-slit reflection and transmission coefficients for a single wave which are not related to the total field structures at the slit edges.
It is also important to know that there is already an extra $\pi$-phase shift in Eq.\,(\ref{e8}) due to opposite signs for the forward-moving wave ($+$) and the backward-moving wave ($-$).
Therefore, for a given incident wavelength $\lambda_0$ in vacuum, the constructive intra-slit dual-wave interference, or the passing state of
a slit, is found to satisfy the condition: $4d=(2m-1)\lambda_0/2\sqrt{\kappa}$ with $m=0,\,\pm 1,\,\pm 2,\,\cdots$, where $\kappa$ is the slit dielectric constant.
Moreover, the condition for the destructive intra-slit dual-wave interference,
or the blocking state of a slit, is given by $4d=2m\lambda_0/2\sqrt{\kappa}$ with $m=\pm 1,\,\pm 2,\,\cdots$. Whether for plane wave or SPP excitation of a single slit, the constructive or destructive dual-wave
interference results agree with the
zero-order transmission coefficient in a single-mode approximation\,\cite{vidal}.
The above interference conditions are not directly derived from the theory described in Sec.\,2. However, these
phenomenological arguments are found to explain the numerical results in this paper pretty well.
\medskip

\begin{figure}[htbp]
\centerline{\includegraphics[width=.8\columnwidth]{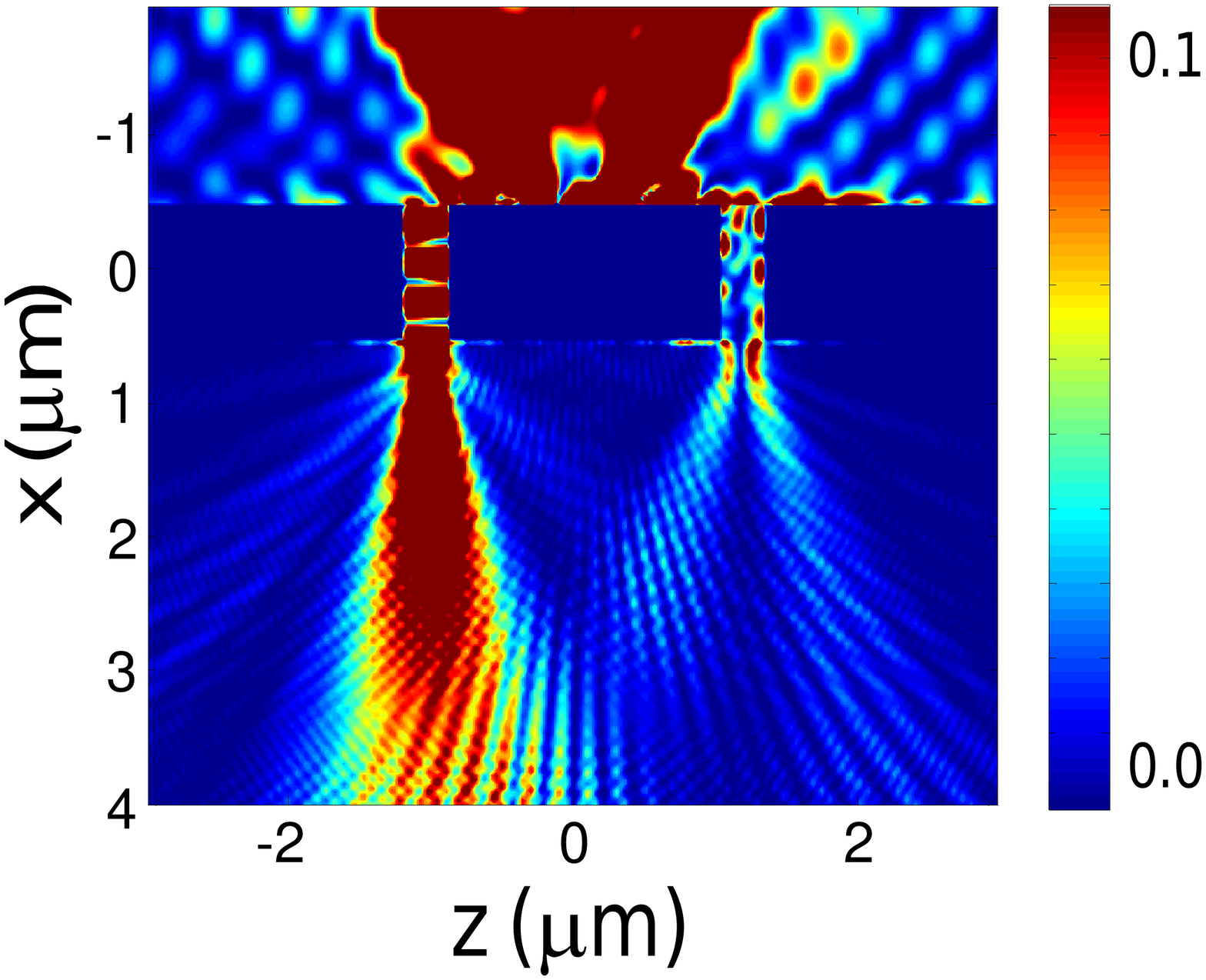}}
\caption{(Color online) Contour plot of $|H_y(x,\,z)|^2$ for $p$-polarization normal incidence (from upper surface) with $\theta_0=0^{\rm o}$, where $n_L=1$ and $n_R=4.5$. In our calculations,
we set the parameters as follows: $\ell_{-1}=\ell_0=\ell_1=\zeta/4$, $\sqrt{\kappa_{-1}}=1$, $\sqrt{\kappa_0}=1+30\,i$, $\sqrt{\kappa_1}=2$,
$z_{-1}/\zeta=-1.75$, $z_0=z_G=0$, $z_1/\zeta=2$, $d=(7/8)\,\zeta$, $g=6\ell_0$, and $\lambda_0=\zeta$, where $\zeta=0.588\,\mu$m.}
\label{f2}
\end{figure}

By employing the mechanism for the intra-slit dual-wave interference discussed above,
Fig.\,\ref{f2} simultaneously displays the passing state of the left slit as well as the blocking state of the right slit at $\lambda_0=0.588\,\mu$m,
where the middle slit is used for a local front-side SP excitation, as seen from Eq.\,(\ref{e6}), and is filled with a wide-band attenuator to prevent light from leaking into Region III.
When the slit depth $2d$ becomes larger than $\lambda_0$ and the slit width $2\ell_j$ is only half of $\lambda_0$, all the slit modes for $p$ polarization become strongly attenuated
except for the lowest symmetric one which has a uniform distribution in the transverse ($z$) direction of a slit. For this lowest symmetric mode, the dual-wave constructive-interference
condition for the left-slit is satisfied due to $4d=7\lambda_0/2\sqrt{\kappa_{-1}}$ and $\sqrt{\kappa_{-1}}=1$. We further find that the dual-wave destructive-interference
condition for the right-slit is met at the same time due to $4d=14\lambda_0/2\sqrt{\kappa_{1}}$ and $\sqrt{\kappa_{1}}=2$.
This fully explains the observed left-slit passing state as well as the right-slit blocking state in Fig.\,\ref{f2}.
\medskip

\begin{figure}[htbp]
\centerline{\includegraphics[width=.8\columnwidth]{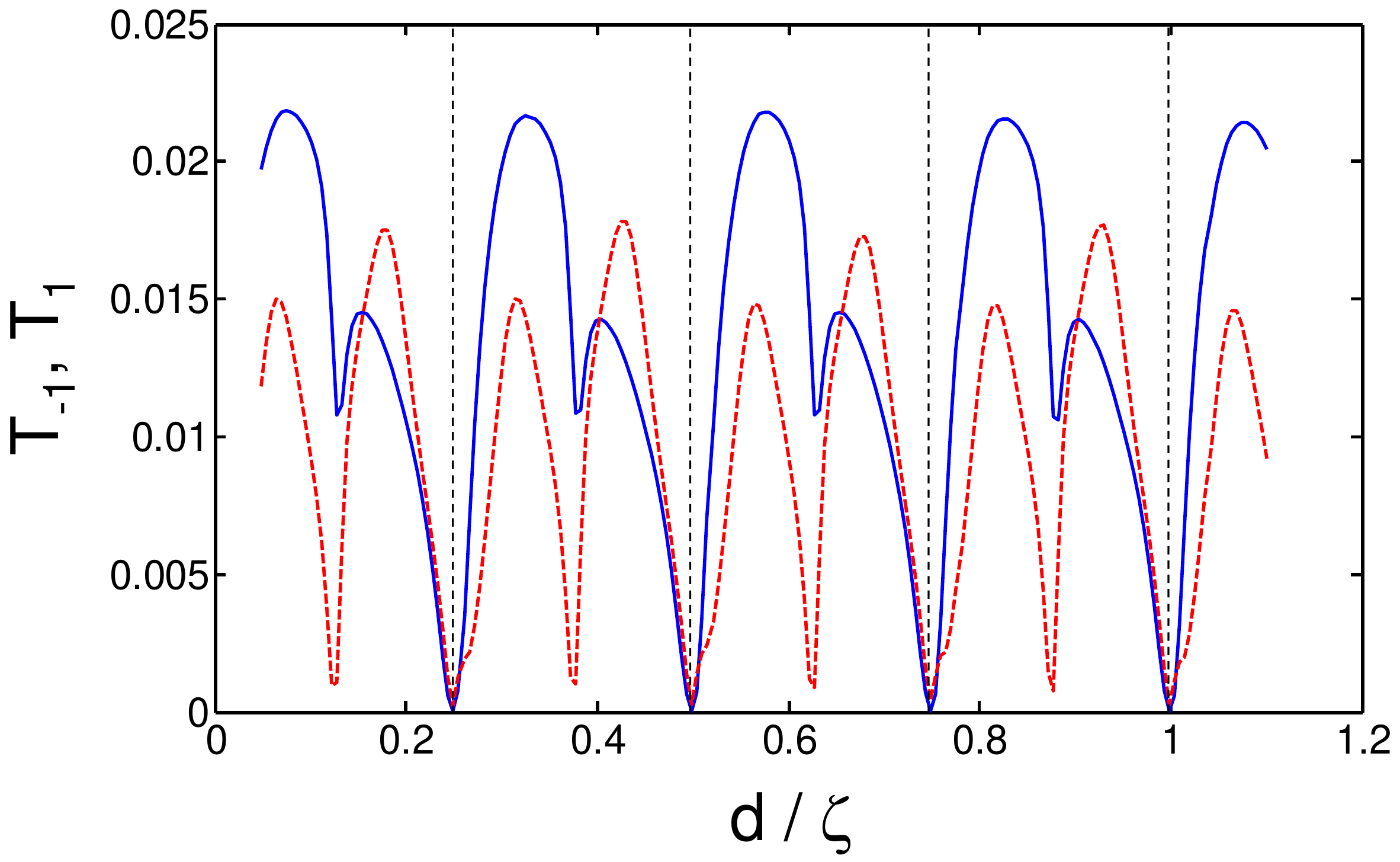}}
\caption{(Color online) Plot for calculated $T_j=\left(1/2\ell_j\right)\int\limits_{-\ell_j}^{\ell_j} dz\, |H_y(d+\delta,\,z-z_j)|^2$ for $j=-1$ (blue solid curve)
and $1$ (red dashed curve) as a function of slit depth $d/\zeta$
for $p$-polarization normal incidence as in Fig.\,\ref{f2}, where $n_L=1$, $n_R=4.5$, $\delta=\zeta/2$ and the vertical black dashed lines indicate
the positions determined by $d/\zeta=2/8,\,4/8,\,6/8,\,8/8$. In our calculations,
we set the parameters as follows: $\ell_{-1}=\ell_0=\ell_1=\zeta/6$, $\sqrt{\kappa_{-1}}=1$, $\sqrt{\kappa_0}=1+30\,i$, $\sqrt{\kappa_1}=2$,
$z_{-1}/\zeta=-1.75$, $z_0=z_G=0$, $z_1/\zeta=2$, $g=6\ell_0$, and $\lambda_0=\zeta$, where $\zeta=0.588\,\mu$m.}
\label{f3}
\end{figure}

In order to get a complete picture about the intra-slit dual-wave interference
after the propagation of an SP wave locally excited at the middle
slit, we present in Fig.\,\ref{f3} the averaged ratio of
$|H_y(x,\,z)|^2$ over a slit at $\lambda_0=0.588\,\mu$m as a function of
slit depth $d$ for two observation slits at a distance $\delta$ slightly away from
the backside ($z=d$) of the metal film. For the left slit at $z=z_{-1}$ (blue solid curve), we find from Fig.\,\ref{f3} that there exist four minima at $d/\lambda_0=2/8$, $4/8$, $6/8$ and $8/8$,
which agree with the dual-wave destructive-interference condition, i.e., $4d=2m\lambda_0/2\sqrt{\kappa_{-1}}$ with $m=1,\,2,\,3,\,4$ and $\sqrt{\kappa_{-1}}=1$.
Similarly, for the right slit at $z=z_1$ (red dashed curve), we find eight minima at $d/\lambda_0=2/16$, $4/16,\,\cdots$, $14/16$ and $16/16$, which also agree with the
right-slit dual-wave destructive-interference condition, i.e., $4d=2m\lambda_0/2\sqrt{\kappa_1}$ with $m=1,\,2,\,\cdots,\,7,\,8$ and $\sqrt{\kappa_{1}}=2$.
In addition, a maximum is always seen between two adjacent minima for the right slit, which meets the dual-wave constructive-interference relation, i.e.,
$4d=(2m-1)\lambda_0/2\sqrt{\kappa_{1}}$ with $m=1,\,2,\,\cdots,\,7,\,8$. However, for the left slit, instead of four maxima as expected,
we only see four dips at $d/\lambda_0=1/8$, $3/8$, $5/8$ and $7/8$ due to the effect of an inter-slit dual-wave interference explained below.
We note that the slit widths in Figs.\,\ref{f2} and \ref{f3} are different, but it will not change the peak and valley positions for the dominant lowest symmetric slit mode.
\medskip

\begin{figure}[htbp]
\centerline{\includegraphics[width=.4\columnwidth]{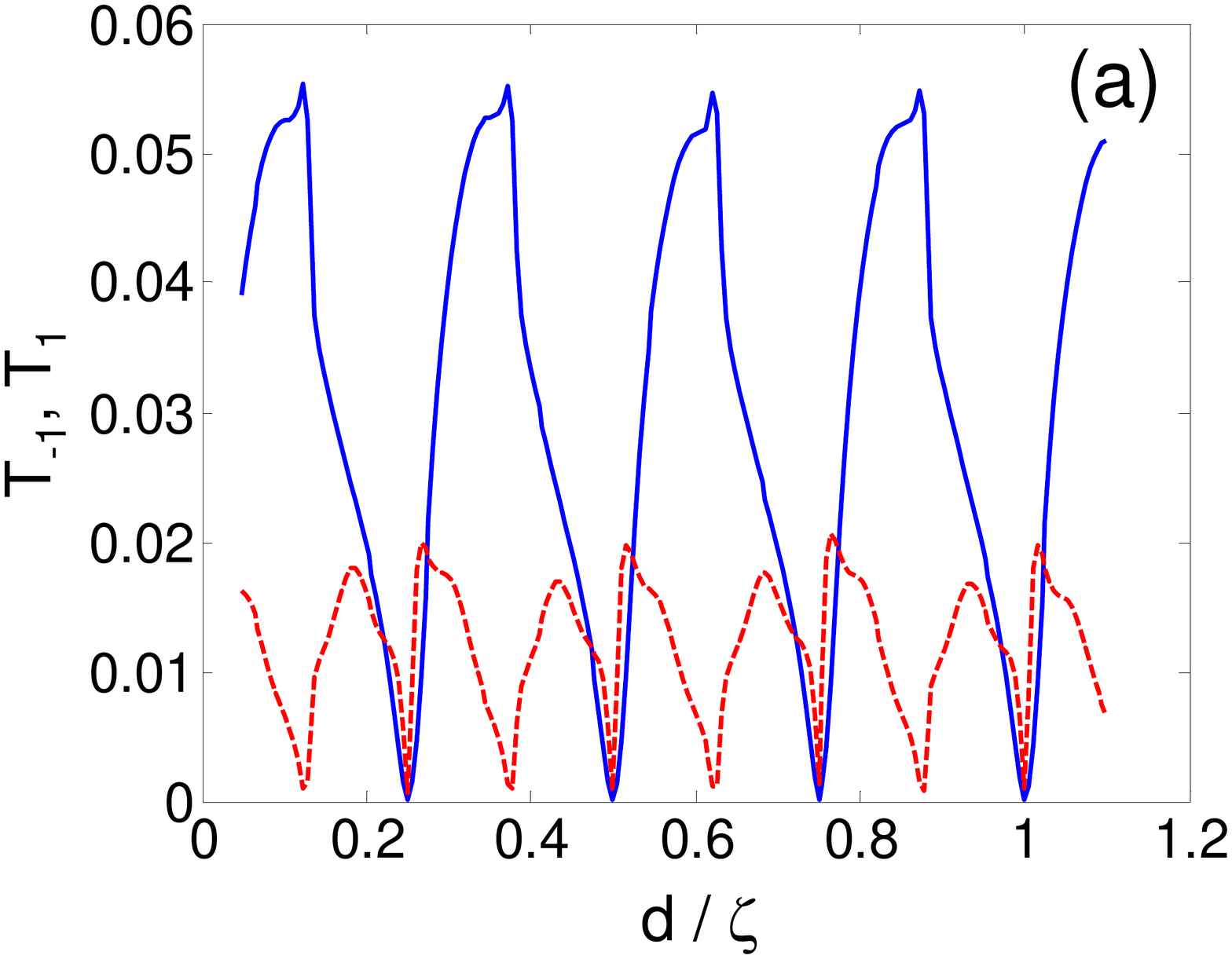}}
\centerline{\includegraphics[width=.4\columnwidth]{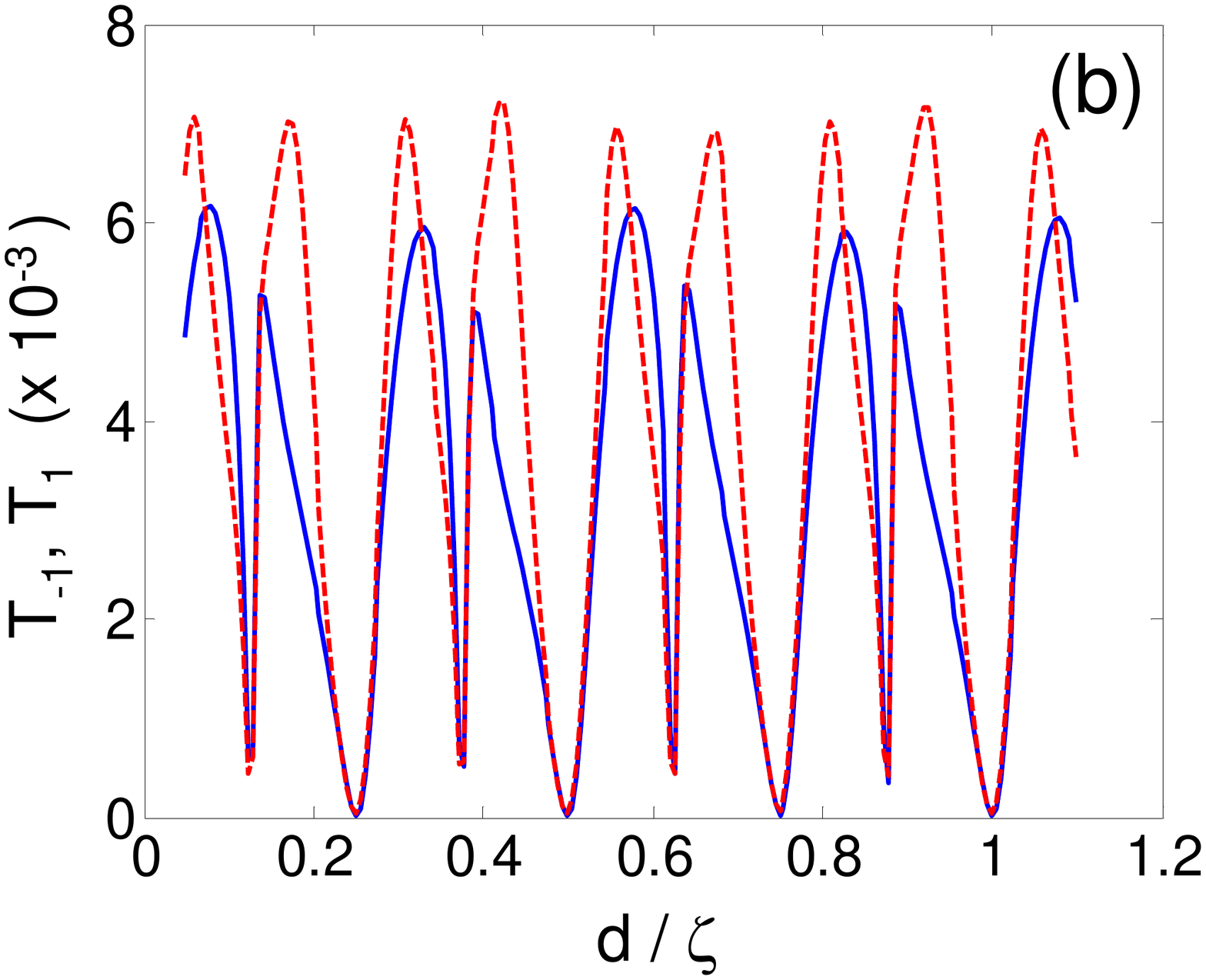}}
\caption{(Color online) Plot for calculated $T_j=\left(1/2\ell_j\right)\int\limits_{-\ell_j}^{\ell_j} dz\, |H_y(d+\delta,\,z-z_j)|^2$ for $j=-1$ (blue solid curve)
and $1$ (red dashed curve) as a function of slit depth $d/\zeta$
for $p$-polarization normal incidence as in Fig.\,\ref{f2}, where $n_L=1$, $n_R=4.5$, $\delta=\zeta/2$. In our calculations,
we set the parameters as follows: $\ell_{-1}=\ell_0=\ell_1=\zeta/6$, $\sqrt{\kappa_{-1}}=1$, $\sqrt{\kappa_0}=1+30\,i$, $\sqrt{\kappa_1}=2$,
$z_0=z_G=0$, $z_1/\zeta=2$, $g=6\ell_0$, and $\lambda_0=\zeta$, where $\zeta=0.588\,\mu$m. Here, we chose $z_{-1}/\zeta=-1.5$ [in (a)] and $z_{-1}/\zeta=-2$ [in (b)], respectively.}
\label{f4}
\end{figure}

It is important to mention that the inter-slit distance in Fig.\,\ref{f3} is set to be $(z_1-z_{-1})/\lambda_0=3.75$, which is a quarter wavelength smaller than a multiple of the wavelength $\lambda_0$.
In order to understand the inter-slit dual-wave interference effect, we show in Fig.\,\ref{f4}
the slit-averaged ratio of $|H_y(x,\,z)|^2$ at $\lambda_0=0.588\,\mu$m as a function of $d$ for $z_{-1}=-1.5\,\lambda_0$ in Fig.\,\ref{f4}(a)
and $z_{-1}=-2\lambda_0$ in Fig.\,\ref{f4}(b) with $z_1=2\lambda_0$ fixed.
When $d/\lambda_0=1/8$, $3/8$, $5/8$ and $7/8$, from Fig.\,\ref{f3} we already know that the contributions from the exit-edge reflected forward-moving waves
are out of phase at the left and right slits since the former is in a
passing state while the latter is in a blocking state.
Therefore, for the case with $(z_1-z_{-1})/\lambda_0=3.5$ in Fig.\,\ref{f4}(a),
we expect a constructive inter-slit dual-wave interference to occur at the left slit (blue solid curve), which can be verified by noticing the change of a dip in Fig.\,\ref{f3} into a maximum in Fig.\,\ref{f4}(a).
However, when $(z_1-z_{-1})/\lambda_0=4$ as shown in Fig.\,\ref{f4}(b), the dip in Fig.\,\ref{f3} changes into a full minimum (blue solid curve) as a consequence of the destructive inter-slit dual-wave interference.
\medskip

\begin{figure}[htbp]
\centerline{\includegraphics[width=.8\columnwidth]{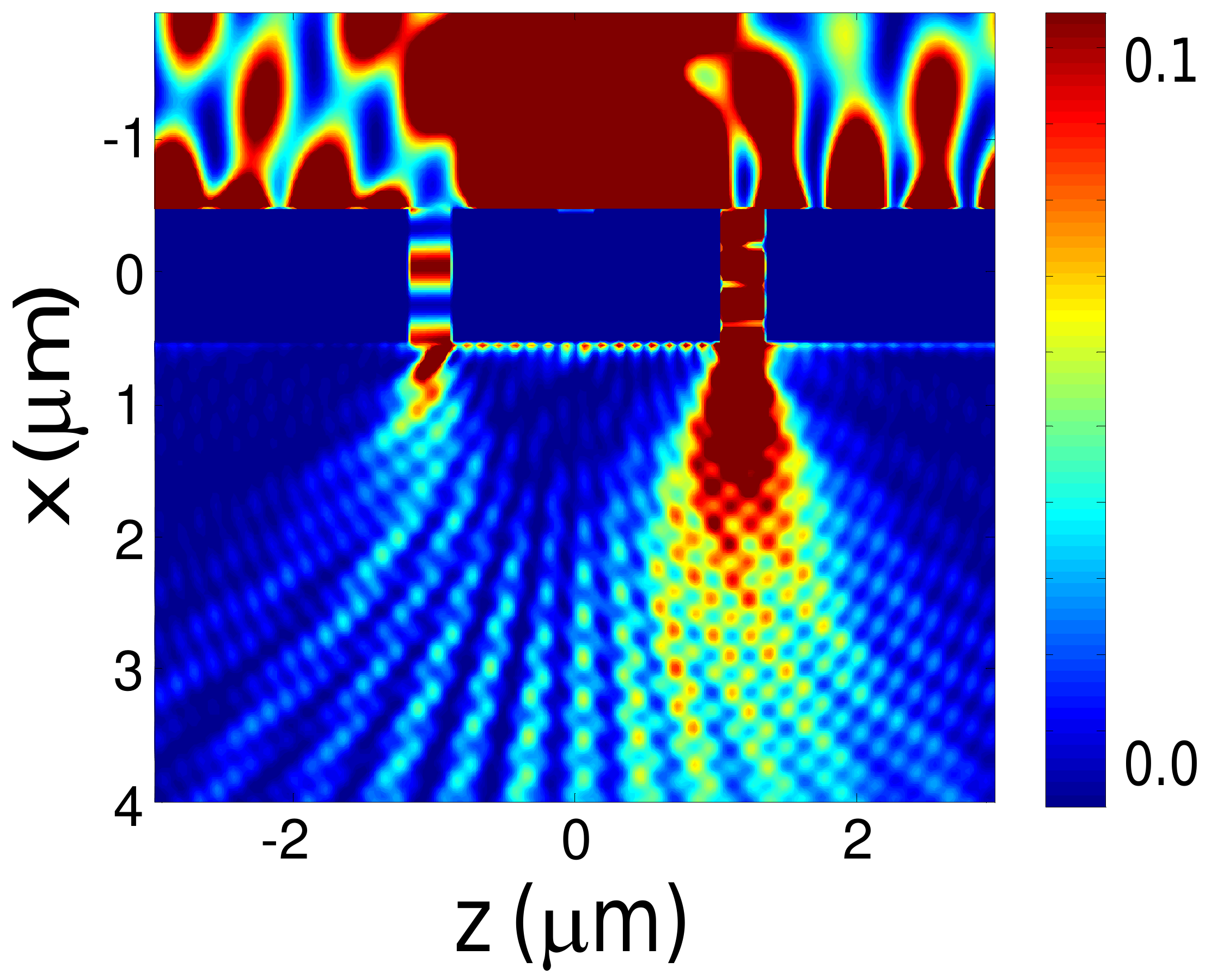}}
\caption{(Color online) Contour plot of $|H_y(x,\,z)|^2$ for $p$-polarization normal incidence (from upper surface) with $\theta_0=0^{\rm o}$, where $n_L=1$ and $n_R=4.5$.
In our calculations,
we set the parameters as follows: $\ell_{-1}=\ell_0=\ell_1=\zeta/4$, $\sqrt{\kappa_{-1}}=1$, $\sqrt{\kappa_0}=1+30\,i$, $\sqrt{\kappa_1}=2$,
$z_{-1}/\zeta=-1.75$, $z_0=z_G=0$, $z_1/\zeta=2$, $d=(7/8)\,\zeta$, $g=6\ell_0$, and $\lambda_0=2\zeta$, where $\zeta=0.588\,\mu$m.}
\label{f5}
\end{figure}

As a complementary result to the left-slit passing state at a particular wavelength $\lambda_0=0.58\,\mu$m in Fig.\,\ref{f2}, we present another contour plot for the EM field distribution at
double that wavelength $\lambda_0=1.176\,\mu$m in Fig.\,\ref{f5}, where the right-slit is in the passing state in this case. Here, the constructive intra-slit dual-wave interference condition
$4d=7\lambda_0/2\sqrt{\kappa_1}$ has been met for the right slit. However, the left slit is neither in the passing state nor in the blocking state at $\lambda_0=1.176\,\mu$m, which is seen in an intermediate state
between these two with $4d=(7/2)\lambda_0/2\sqrt{\kappa_{-1}}$.
\medskip

\begin{figure}[htbp]
\centerline{\includegraphics[width=.8\columnwidth]{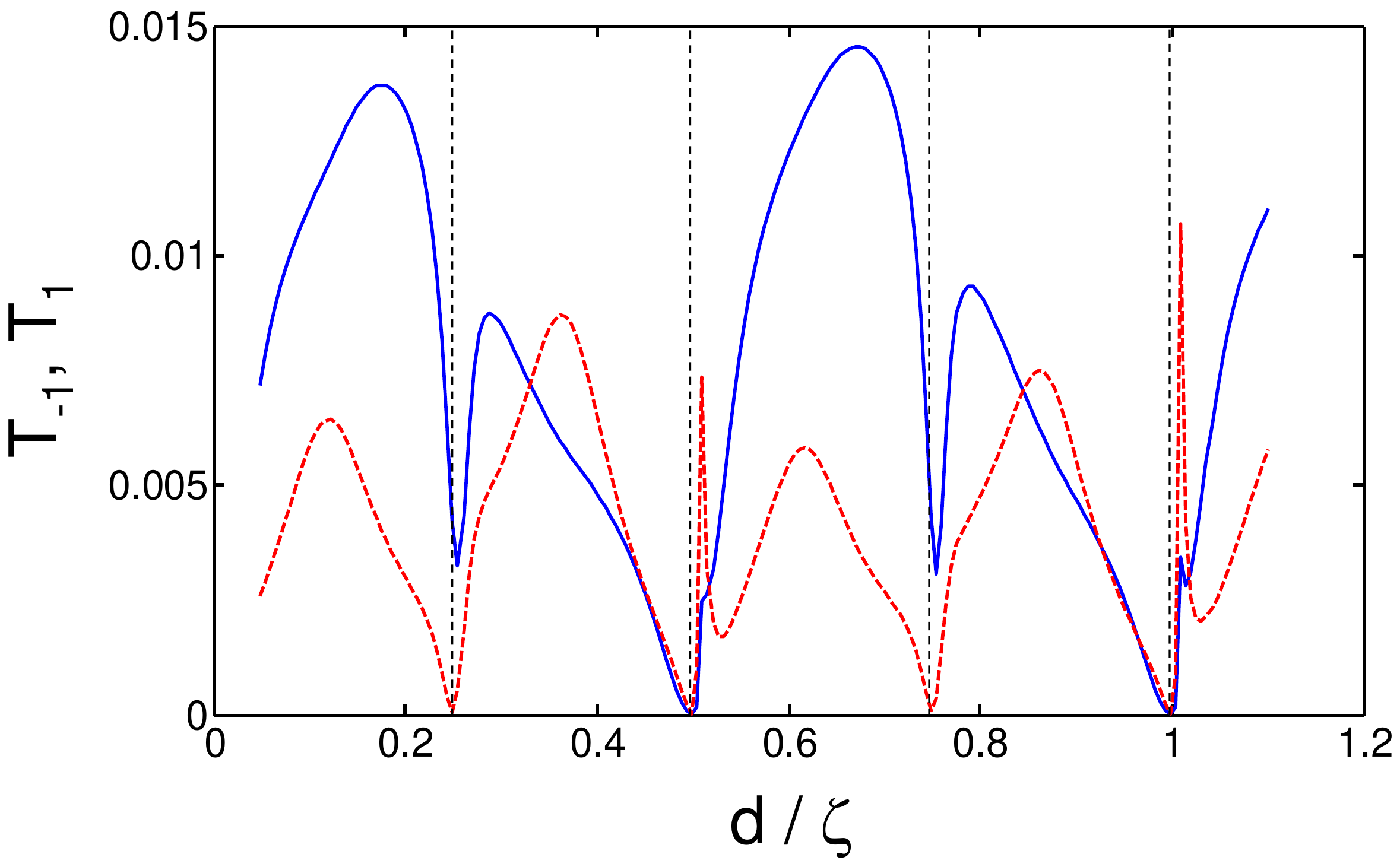}}
\caption{(Color online) Plot for calculated $T_j=\left(1/2\ell_j\right)\int\limits_{-\ell_j}^{\ell_j} dz\, |H_y(d+\delta,\,z-z_j)|^2$ for $j=-1$ (blue solid curve)
and $1$ (red dashed curve) as a function of slit depth $d/\zeta$
for $p$-polarization normal incidence as in Fig.\,\ref{f5}, where $n_L=1$, $n_R=4.5$, $\delta=\zeta/2$ and  the vertical black dashed lines indicate
the positions determined by $d/\zeta=2/8,\,4/8,\,6/8,\,8/8$.. In our calculations,
we set the parameters as follows: $\ell_{-1}=\ell_0=\ell_1=\zeta/6$, $\sqrt{\kappa_{-1}}=1$, $\sqrt{\kappa_0}=1+30\,i$, $\sqrt{\kappa_1}=2$,
$z_{-1}/\zeta=-1.75$, $z_0=z_G=0$, $z_1/\zeta=2$, $g=6\ell_0$, and $\lambda_0=2\zeta$, where $\zeta=0.588\,\mu$m.}
\label{f6}
\end{figure}

To get a complete picture about the intra-slit dual-wave interference at $\lambda_0=1.176\,\mu$m, we show the slit-averaged ratio of $|H_y(x,\,z)|^2$
at $\lambda_0=1.176\,\mu$m in Fig.\,\ref{f6} as a function of $d$. Indeed, we find from this figure that there exists a set of minima for the right slit
at $2d/\lambda_0=2/8$, $4/8$, $6/8$ and $8/8$ with $\sqrt{\kappa_1}=2$, as well as a group of maxima at $2d/\lambda_0=1/8$, $3/8$, $5/8$ and $7/8$. We also notice here that dips, instead of maxima,
still show up for the left slit at $2d/\lambda_0=2/8$ and $6/8$ with $\sqrt{\kappa_{-1}}=1$ due to incomplete inter-slit dual-wave constructive interference as discussed in connection with Fig.\,\ref{f4}.
Interestingly, we also find very narrow peaks right above two of the four blocking states of the right slit, and these sharp peaks get stronger with increased slit depth. The occurrence of two extremely sharp
peaks in this figure can be qualitatively attributed to the result of the Fabry-P\'erot sole-wave interference between the entry and exit edges of the right slit with a finite value for the average
reflection coefficient (or a large finesse).
The peak positions from a constructive Fabry-P\'erot interference coincidentally overlaps with two of the four slit blocking states.
Although the slit widths in Figs.\,\ref{f5} and \ref{f6} are different, it will not change the peak and valley positions for the dominant lowest symmetric slit mode.
\medskip

\begin{figure}[htbp]
\centerline{\includegraphics[width=.8\columnwidth]{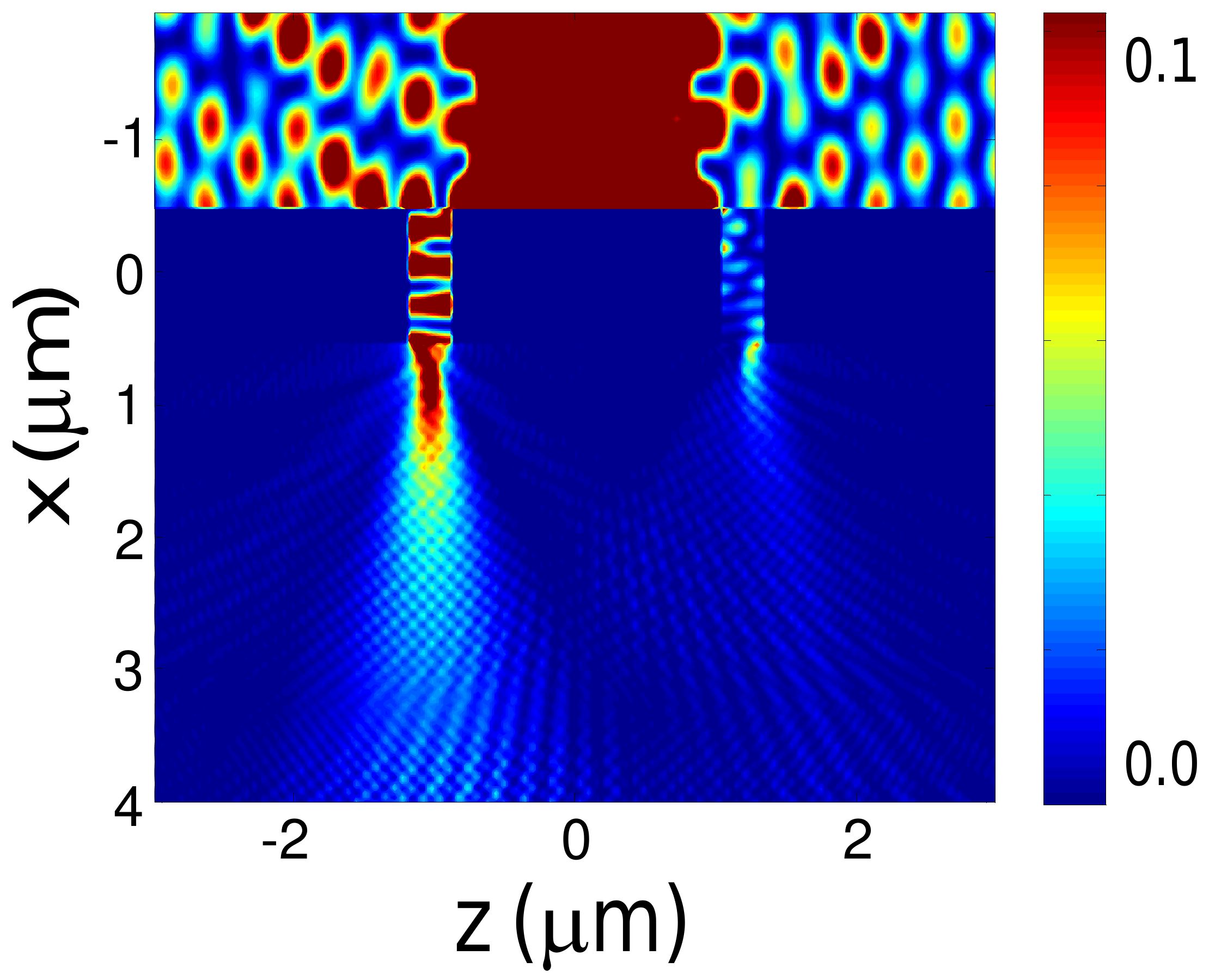}}
\caption{(Color online) Contour plot of $|H_y(x,\,z)|^2$ for $p$-polarization normal incidence (from upper surface) with $\theta_0=0^{\rm o}$, where $n_L=1$ and $n_R=4.5$. In our calculations,
we set the parameters as follows: $\ell_{-1}=\ell_0=\ell_1=\zeta/4$, $\sqrt{\kappa_{-1}}=1$, $\sqrt{\kappa_0}=1+30\,i$, $\sqrt{\kappa_1}=2$,
$z_{-1}/\zeta=-1.75$, $z_0=z_G=0$, $z_1/\zeta=2$, $d=(7/8)\,\zeta$, $g=6\ell_0$, and $\lambda_0=\zeta$, where $\zeta=0.588\,\mu$m. In this case, we set $\eta_L=\eta_R=0$ for a perfect electric conductor.}
\label{f7}
\end{figure}

The current paper deals with a real metallic film by imposing the
SIBC in Eqs.\,(\ref{e1}) and (\ref{e2}) for a finite metal conductivity. This facilitates the
propagation of the SP polariton wave, which is excited
at the middle slit, to two neighboring side slits. Consequently, for a
PEC with $\eta_L=\eta_R=0$ we expect the passing state of the left slit in Fig.\,\ref{f2}
will become much weaker, which can be clearly seen from the
calculated transmitted near-field distribution in Fig.\,\ref{f7}.
In the case of a PEC, no SP polariton wave is excited on the backside of the metal film, and the reflection from the illuminated
middle slit on the front-side of the film becomes very collimated in the near-field region.
Moreover, the maximum intensity in Fig.\,\ref{f2} occurs farther from the film than it does in Fig.\,\ref{f7}, and
the angular distribution of the intensity of the transmitted field is broader in Fig.\,\ref{f2} than it is in Fig.\,\ref{f7}.
There exists a major difference in the coupling between slits on the interface of a PEC or a real metal. For the former, only radiative modes out of the metal plane
can contribute, while both surface-plasmon and radiative modes will contribute to the latter.
\medskip

\begin{figure}[htbp]
\centerline{\includegraphics[width=.8\columnwidth]{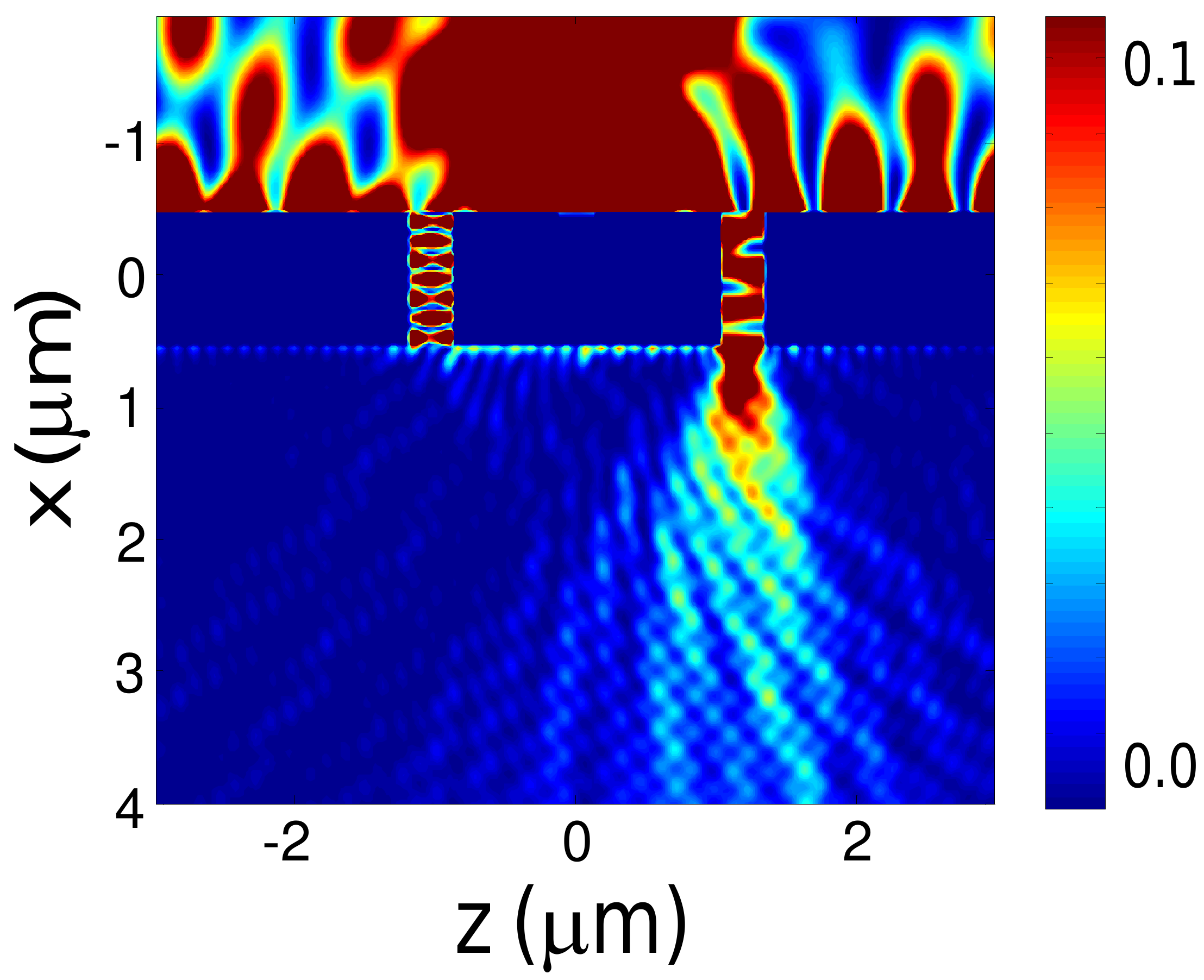}}
\caption{(Color online) Contour plot of $|H_y(x,\,z)|^2$ for $p$-polarization normal incidence (from upper surface) with $\theta_0=0^{\rm o}$, where $n_L=1$ and $n_R=4.5$. In our calculations,
we set the parameters as follows: $\ell_{-1}=\ell_0=\ell_1=\zeta/4$, $\sqrt{\kappa_{-1}}=4$, $\sqrt{\kappa_0}=1+30\,i$, $\sqrt{\kappa_1}=2$,
$z_{-1}/\zeta=-1.75$, $z_0=z_G=0$, $z_1/\zeta=2$, $d=(7/8)\,\zeta$, $g=6\ell_0$, and $\lambda_0=2\zeta$, where $\zeta=0.588\,\mu$m.}
\label{f8}
\end{figure}

Finally, we know from Fig.\,\ref{f5} that the left slit is in
an intermediate state although the right slit is in a passing
state at $\lambda_0=1.176\,\mu$m. The ideal
situation is that the left slit could be forced into a blocking state at
this wavelength. This goal can be reached if
the filled dielectric medium in the left slit can be tuned from $\sqrt{\kappa_{-1}}=1$
to $\sqrt{\kappa_{-1}}=4$, which has been simulated by the calculated transmitted
near-field distribution displayed in Fig.\,\ref{f8}.

\section{Conclusions}
\label{sec4}

In conclusion, motivated by the previous Green's function formalism\,\cite{Baumeier,Wellems}, we have derived a scattering-wave theory to study the wavelength-dependent detection of
surface-plasmon mediated light-beam splitting into two by a triple-slit structure perforated by a gold film and filled with different slit materials. This can be viewed
as a new addition to the previously demonstrated
longitudinal color-dependent light focusing using a finite groove array with various groove widths in a parabolic pattern.
For a specifically chosen slit depth, filled slits dielectric material, and inter-slit distance in the deep sub-wavelength regime,
we have found that only one of the two side observation slits is in a passing state for a particular wavelength, but the
other blocked slit switches to a passing state at double that wavelength. In this sense, surface-plasmon mediated light-beam splitting becomes wavelength sensitive,
and a single light-beam incidence with two wavelengths can be separated along the transverse direction parallel to the array.
This provides us with a direct optical reading in the near-field region based on a non-spectroscopic technique.\\
\medskip

\clearpage
\noindent
{\bf Acknowledgments}\\

We would like to thank the Air Force Office of Scientific Research
(AFOSR) for its support. We would also like to thank helpful suggestions as well as a critical reading of the paper by Prof. A. A. Maradudin.
\\
\medskip

\noindent
{\bf Appendix A}\\

The overlap integrals initially introduced in Eqs.\,(\ref{e13}) and (\ref{e14}) are defined as follows:

\begin{eqnarray}
Q_{sn}^j(\beta)&=&
\frac{1}{\ell_j}\,\int\limits_{-\ell_j}^{\ell_j} dz\,\cos(\xi_{sn}^jz)\,\cos(\beta z)
\nonumber\\
&=&{\rm sinc}[(\beta-\xi_{sn}^j)\,\ell_j]+{\rm sinc}[(\beta+\xi_{sn}^j)\,\ell_j]\ ,
\label{a1}
\end{eqnarray}

\begin{eqnarray}
Q_{an}^j(\beta)&=&
\frac{1}{\ell_j}\,\int\limits_{-\ell_j}^{\ell_j} dz\,\sin(\xi_{an}^jz)\,\sin(\beta z)
\nonumber\\
&=&{\rm sinc}[(\beta-\xi_{an}^j)\,\ell_j]-{\rm sinc}[(\beta+\xi_{an}^j)\,\ell_j]\ ,
\label{a2}
\end{eqnarray}
and ${\rm sinc}(x)\equiv\sin x/x$.
\\
\medskip

\noindent
{\bf Appendix B}\\

The coupling matrices initially introduced in Eqs.\,(\ref{e13}), (\ref{e14}),
(\ref{e17}) and (\ref{e18}) are defined as

\begin{eqnarray}
&&P_s(\beta,\,\beta^\prime)=\int\limits_{-\infty}^{-|z_{-N}-\ell_{-N}|} \cos(\beta^\prime z)\,\cos(\beta z)\,dz + \int\limits_{z_{N}+\ell_{N}}^{\infty} \cos(\beta^\prime z)\,\cos(\beta z)\,dz
\nonumber\\
&=&\pi\left[\delta(\beta-\beta^\prime)+\delta(\beta+\beta^\prime)\right]
\nonumber\\
&-&\frac{|z_{-N}-\ell_{-N}|}{2}\,\left\{{\rm sinc}[(\beta+\beta^\prime)|z_{-N}-\ell_{-N}|]+{\rm sinc}[(\beta-\beta^\prime)|z_{-N}-\ell_{-N}|]\right\}
\nonumber\\
&-&\frac{z_{N}+\ell_{N}}{2}\,\left\{{\rm sinc}[(\beta+\beta^\prime)(z_{N}+\ell_{N})]+{\rm sinc}[(\beta-\beta^\prime)(z_{N}+\ell_{N})]\right\}\ ,
\label{b1}
\end{eqnarray}

\begin{eqnarray}
&&P_a(\beta,\,\beta^\prime)=\int\limits_{-\infty}^{-|z_{-N}-\ell_{-N}|} \sin(\beta^\prime z)\,\sin(\beta z)\,dz + \int\limits_{z_{N}+\ell_{N}}^{\infty} \sin(\beta^\prime z)\,\sin(\beta z)\,dz
\nonumber\\
&=&\pi\left[\delta(\beta-\beta^\prime)-\delta(\beta+\beta^\prime)\right]
\nonumber\\
&+&\frac{|z_{-N}-\ell_{-N}|}{2}\,\left\{{\rm sinc}[(\beta+\beta^\prime)|z_{-N}-\ell_{-N}|]-{\rm sinc}[(\beta-\beta^\prime)|z_{-N}-\ell_{-N}|]\right\}
\nonumber\\
&+&\frac{z_{N}+\ell_{N}}{2}\,\left\{{\rm sinc}[(\beta+\beta^\prime)(z_{N}+\ell_{N})]-{\rm sinc}[(\beta-\beta^\prime)(z_{N}+\ell_{N})]\right\}\ ,
\label{b2}
\end{eqnarray}

\begin{eqnarray}
&&P_c(\beta,\,\beta^\prime)=\int\limits_{-\infty}^{-|z_{-N}-\ell_{-N}|} \sin(\beta^\prime z)\,\cos(\beta z)\,dz + \int\limits_{z_{N}+\ell_{N}}^{\infty} \sin(\beta^\prime z)\,\cos(\beta z)\,dz
\nonumber\\
&-&\frac{1}{2}\,\left\{\frac{\cos[(\beta+\beta^\prime)|z_{-N}-\ell_{-N}|]}{\beta+\beta^\prime}-\frac{\cos[(\beta-\beta^\prime)|z_{-N}-\ell_{-N}|]}{\beta-\beta^\prime}\right\}
\nonumber\\
&+&\frac{1}{2}\,\left\{\frac{\cos[(\beta+\beta^\prime)(z_{N}+\ell_{N})]}{\beta+\beta^\prime}-\frac{\cos[(\beta-\beta^\prime)(z_{N}+\ell_{N})]}{\beta-\beta^\prime}\right\}\ ,
\label{b3}
\end{eqnarray}

\begin{eqnarray}
&&W_s(\beta,\,\beta^\prime)=\sum\limits_{j=-N}^{N-1}\int\limits_{z_j+ \ell_j}^{z_{j+1}-\ell_{j+1}} \cos(\beta^\prime z)\,\cos(\beta z)\,dz
\nonumber\\
&=&\frac{1}{2}\,\sum\limits_{j=-N}^{N-1}\left\{(z_{j+1}-\ell_{j+1})\,\left({\rm sinc}[(\beta^\prime-\beta)(z_{j+1}-\ell_{j+1})]\right.\right.
\nonumber\\
&+&\left.{\rm sinc}[(\beta^\prime+\beta)(z_{j+1}-\ell_{j+1})]\right)
\nonumber\\
&-&\left.(z_{j}+\ell_{j})\,\left({\rm sinc}[(\beta^\prime+\beta)(z_{j}+\ell_{j})]+{\rm sinc}[(\beta^\prime-\beta)(z_{j}+\ell_{j})]\right)\right\}\ ,
\label{b4}
\end{eqnarray}

\begin{eqnarray}
&&W_a(\beta,\,\beta^\prime)=\sum\limits_{j=-N}^{N-1}\int\limits_{z_j+ \ell_j}^{z_{j+1}-\ell_{j+1}} \sin(\beta^\prime z)\,\sin(\beta z)\,dz
\nonumber\\
&=&\frac{1}{2}\,\sum\limits_{j=-N}^{N-1}\left\{(z_{j+1}-\ell_{j+1})\,\left({\rm sinc}[(\beta^\prime-\beta)(z_{j+1}-\ell_{j+1})]\right.\right.
\nonumber\\
&-&\left.{\rm sinc}[(\beta^\prime+\beta)(z_{j+1}-\ell_{j+1})]\right)
\nonumber\\
&+&\left.(z_{j}+\ell_{j})\,\left({\rm sinc}[(\beta^\prime+\beta)(z_{j}+\ell_{j})]-{\rm sinc}[(\beta^\prime-\beta)(z_{j}+\ell_{j})]\right)\right\}\ ,
\label{b5}
\end{eqnarray}

\begin{eqnarray}
&&W_c(\beta,\,\beta^\prime)=\sum\limits_{j=-N}^{N-1}\int\limits_{z_j+ \ell_j}^{z_{j+1}-\ell_{j+1}} \sin(\beta^\prime z)\,\cos(\beta z)\,dz
\nonumber\\
&=&\frac{1}{2}\,\sum\limits_{j=-N}^{N-1}\left\{\frac{\cos[(\beta^\prime-\beta)(z_{j}+\ell_{j})]}{\beta^\prime-\beta}
+\frac{\cos[(\beta^\prime+\beta)(z_{j}+\ell_{j})]}{\beta^\prime+\beta}\right.
\nonumber\\
&-&\left.\frac{\cos[(\beta^\prime-\beta)(z_{j+1}-\ell_{j+1})]}{\beta^\prime-\beta}-\frac{\cos[(\beta^\prime+\beta)(z_{j+1}-\ell_{j+1})]}{\beta^\prime+\beta}\right\}\ .
\label{b6}
\end{eqnarray}
\\
\medskip

\noindent
{\bf Appendix C}\\

We have introduced in Eq.\,(\ref{e29}) the following amplitudes

\begin{eqnarray}
&&Y^{(1)}_{jn}=\frac{\sigma_{sn}^j \epsilon_L}{\chi_n
\kappa_j}\,\int\limits_{0}^{\infty}
\frac{d\beta}{k_1(\beta)}\,Q_{sn}^j(\beta)
\nonumber\\
&\times&\left\{\left[G_s(\beta)-A_s(\beta))\right]\,\cos(\beta z_j) +i\left[G_a(\beta)-A_a(\beta)\right]\,\sin(\beta z_j)\right\}\ ,
\label{c1}
\end{eqnarray}

\begin{equation}
Y^{(2)}_{jn}=\frac{\sigma_{sn}^j \epsilon_R}{\chi_n
\kappa_j}\,\int\limits_{0}^{\infty}
\frac{d\beta}{k_2(\beta)}\,Q_{sn}^j(\beta)\,
\left[B_s(\beta)\,\cos(\beta z_j) + iB_a(\beta)\,\sin(\beta
z_j)\right]\ ,
\label{c2}
\end{equation}

\begin{eqnarray}
&&X^{(1)}_{jn}=\frac{\sigma_{an}^j \epsilon_L}{\chi_n
\kappa_j}\,\int\limits_{0}^{\infty}
\frac{d\beta}{k_1(\beta)}\,Q_{an}^j(\beta)
\nonumber\\
&\times&\left\{\left[G_a(\beta)-A_a(\beta)\right]\,\cos(\beta z_j)
+i\left[G_s(\beta)-A_s(\beta)\right]\,\sin(\beta z_j)\right\}\ ,
\label{c3}
\end{eqnarray}

\begin{equation}
X^{(2)}_{jn}=\frac{\sigma_{an}^j \epsilon_R}{\chi_n
\kappa_j}\,\int\limits_{0}^{\infty}
\frac{d\beta}{k_2(\beta)}\,Q_{an}^j(\beta)\,
\left[B_a(\beta)\,\cos(\beta z_j) + i B_s(\beta)\,\sin(\beta
z_j)\right]\ . \label{c4}
\end{equation}
\\
\medskip

\noindent
{\bf Appendix D}\\

In Eqs.\,(\ref{e34}) through (\ref{e37}) we have introduced the following matrices

\[
\left[\begin{array}{c}
T_s^{(1)}(\beta,\,\beta^\prime)\\
T_s^{(2)}(\beta,\,\beta^\prime)\\
T_s^{(3)}(\beta,\,\beta^\prime)\\
T_s^{(4)}(\beta,\,\beta^\prime)
\end{array}\right]
=\frac{1}{i\pi}\,\sum_{n,\,j}\,\left[\frac{\sigma_{sn}^j\ell_j}{\chi_n\kappa_j\sin(2\sigma_{sn}^{j}d)}\right]\,Q_{sn}^j(\beta)\,
Q_{sn}^j(\beta^\prime)
\]
\begin{equation}
\times\left[\begin{array}{c}
\epsilon_R\cos(\beta z_j)\cos(\beta^\prime z_j)\\
\epsilon_R\cos(\beta z_j)\sin(\beta^\prime z_j)\\
\epsilon_L\cos(2\sigma_{sn}^{j}d)\,\cos(\beta z_j)\cos(\beta^\prime z_j)\\
\epsilon_L\cos(2\sigma_{sn}^{j}d)\,\cos(\beta z_j)\sin(\beta^\prime
z_j)
\end{array}\right]\ ,
\label{d1}
\end{equation}

\[
\left[\begin{array}{c}
R_s^{(1)}(\beta,\,\beta^\prime)\\
R_s^{(2)}(\beta,\,\beta^\prime)\\
R_s^{(3)}(\beta,\,\beta^\prime)\\
R_s^{(4)}(\beta,\,\beta^\prime)
\end{array}\right]
=\frac{1}{i\pi}\,\sum_{n,\,j}\,\left[\frac{\sigma_{an}^j\ell_j}{\chi_n\kappa_j\sin(2\sigma_{an}^{j}d)}\right]\,Q_{an}^j(\beta)\,
Q_{an}^j(\beta^\prime)
\]
\begin{equation}
\times\left[\begin{array}{c}
\epsilon_R\sin(\beta z_j)\cos(\beta^\prime z_j)\\
\epsilon_R\sin(\beta z_j)\sin(\beta^\prime z_j)\\
\epsilon_L\cos(2\sigma_{an}^{j}d)\,\sin(\beta z_j)\cos(\beta^\prime z_j)\\
\epsilon_L\cos(2\sigma_{an}^{j}d)\,\sin(\beta z_j)\sin(\beta^\prime
z_j)
\end{array}\right]\ ,
\label{d2}
\end{equation}

\[
\left[\begin{array}{c}
T_a^{(1)}(\beta,\,\beta^\prime)\\
T_a^{(2)}(\beta,\,\beta^\prime)\\
T_a^{(3)}(\beta,\,\beta^\prime)\\
T_a^{(4)}(\beta,\,\beta^\prime)
\end{array}\right]
=\frac{1}{i\pi}\,\sum_{n,\,j}\,\left[\frac{\sigma_{sn}^j\ell_j}{\chi_n\kappa_j\sin(2\sigma_{sn}^{j}d)}\right]\,Q_{sn}^j(\beta)\,
Q_{sn}^j(\beta^\prime)
\]
\begin{equation}
\times\left[\begin{array}{c}
\epsilon_R\sin(\beta z_j)\cos(\beta^\prime z_j)\\
\epsilon_R\sin(\beta z_j)\sin(\beta^\prime z_j)\\
\epsilon_L\cos(2\sigma_{sn}^{j}d)\,\sin(\beta z_j)\cos(\beta^\prime z_j)\\
\epsilon_L\cos(2\sigma_{sn}^{j}d)\,\sin(\beta z_j)\sin(\beta^\prime
z_j)
\end{array}\right]\ ,
\label{d3}
\end{equation}

\[
\left[\begin{array}{c}
R_a^{(1)}(\beta,\,\beta^\prime)\\
R_a^{(2)}(\beta,\,\beta^\prime)\\
R_a^{(3)}(\beta,\,\beta^\prime)\\
R_a^{(4)}(\beta,\,\beta^\prime)
\end{array}\right]
=\frac{1}{i\pi}\,\sum_{n,\,j}\,\left[\frac{\sigma_{an}^j\ell_j}{\chi_n\kappa_j\sin(2\sigma_{an}^{j}d)}\right]\,Q_{an}^j(\beta)\,
Q_{an}^j(\beta^\prime)
\]
\begin{equation}
\times\left[\begin{array}{c}
\epsilon_R\cos(\beta z_j)\cos(\beta^\prime z_j)\\
\epsilon_R\cos(\beta z_j)\sin(\beta^\prime z_j)\\
\epsilon_L\cos(2\sigma_{an}^{j}d)\,\cos(\beta z_j)\cos(\beta^\prime z_j)\\
\epsilon_L\cos(2\sigma_{an}^{j}d)\,\cos(\beta z_j)\sin(\beta^\prime
z_j)
\end{array}\right]\ ,
\label{d4}
\end{equation}

\[
\left[\begin{array}{c}
N_s^{(1)}(\beta,\,\beta^\prime)\\
N_s^{(2)}(\beta,\,\beta^\prime)\\
N_s^{(3)}(\beta,\,\beta^\prime)\\
N_s^{(4)}(\beta,\,\beta^\prime)
\end{array}\right]
=\frac{1}{i\pi}\,\sum_{n,\,j}\,\left[\frac{\sigma_{sn}^j\ell_j}{\chi_n\kappa_j\sin(2\sigma_{sn}^{j}d)}\right]\,Q_{sn}^j(\beta)\,
Q_{sn}^j(\beta^\prime)
\]
\begin{equation}
\times\left[\begin{array}{c}
\epsilon_R\cos(2\sigma_{sn}^{j}d)\,\cos(\beta z_j)\cos(\beta^\prime z_j)\\
\epsilon_R\cos(2\sigma_{sn}^{j}d)\,\cos(\beta z_j)\sin(\beta^\prime z_j)\\
\epsilon_L\cos(\beta z_j)\cos(\beta^\prime z_j)\\
\epsilon_L\cos(\beta z_j)\sin(\beta^\prime z_j)
\end{array}\right]\ ,
\label{d5}
\end{equation}

\[
\left[\begin{array}{c}
M_s^{(1)}(\beta,\,\beta^\prime)\\
M_s^{(2)}(\beta,\,\beta^\prime)\\
M_s^{(3)}(\beta,\,\beta^\prime)\\
M_s^{(4)}(\beta,\,\beta^\prime)
\end{array}\right]
=\frac{1}{i\pi}\,\sum_{n,\,j}\,\left[\frac{\sigma_{an}^j\ell_j}{\chi_n\kappa_j\sin(2\sigma_{an}^{j}d)}\right]\,Q_{an}^j(\beta)\,
Q_{an}^j(\beta^\prime)
\]
\begin{equation}
\times\left[\begin{array}{c}
\epsilon_R\cos(2\sigma_{an}^{j}d)\,\sin(\beta z_j)\cos(\beta^\prime z_j)\\
\epsilon_R\cos(2\sigma_{an}^{j}d)\,\sin(\beta z_j)\sin(\beta^\prime z_j)\\
\epsilon_L\sin(\beta z_j)\cos(\beta^\prime z_j)\\
\epsilon_L\sin(\beta z_j)\sin(\beta^\prime z_j)
\end{array}\right]\ ,
\label{d6}
\end{equation}

\[
\left[\begin{array}{c}
N_a^{(1)}(\beta,\,\beta^\prime)\\
N_a^{(2)}(\beta,\,\beta^\prime)\\
N_a^{(3)}(\beta,\,\beta^\prime)\\
N_a^{(4)}(\beta,\,\beta^\prime)
\end{array}\right]
=\frac{1}{i\pi}\,\sum_{n,\,j}\,\left[\frac{\sigma_{sn}^j\ell_j}{\chi_n\kappa_j\sin(2\sigma_{sn}^{j}d)}\right]\,Q_{sn}^j(\beta)\,
Q_{sn}^j(\beta^\prime)
\]
\begin{equation}
\times\left[\begin{array}{c}
\epsilon_R\cos(2\sigma_{sn}^{j}d)\,\sin(\beta z_j)\cos(\beta^\prime z_j)\\
\epsilon_R\cos(2\sigma_{sn}^{j}d)\,\sin(\beta z_j)\sin(\beta^\prime z_j)\\
\epsilon_L\sin(\beta z_j)\cos(\beta^\prime z_j)\\
\epsilon_L\sin(\beta z_j)\sin(\beta^\prime z_j)
\end{array}\right]\ ,
\label{d7}
\end{equation}

\[
\left[\begin{array}{c}
M_a^{(1)}(\beta,\,\beta^\prime)\\
M_a^{(2)}(\beta,\,\beta^\prime)\\
M_a^{(3)}(\beta,\,\beta^\prime)\\
M_a^{(4)}(\beta,\,\beta^\prime)
\end{array}\right]
=\frac{1}{i\pi}\,\sum_{n,\,j}\,\left[\frac{\sigma_{an}^j\ell_j}{\chi_n\kappa_j\sin(2\sigma_{an}^{j}d)}\right]\,Q_{an}^j(\beta)\,
Q_{an}^j(\beta^\prime)
\]
\begin{equation}
\times\left[\begin{array}{c}
\epsilon_R\cos(2\sigma_{an}^{j}d)\,\cos(\beta z_j)\cos(\beta^\prime z_j)\\
\epsilon_R\cos(2\sigma_{an}^{j}d)\,\cos(\beta z_j)\sin(\beta^\prime z_j)\\
\epsilon_L\cos(\beta z_j)\cos(\beta^\prime z_j)\\
\epsilon_L\cos(\beta z_j)\sin(\beta^\prime z_j)
\end{array}\right]\ .
\label{d8}
\end{equation}

\end{document}